\documentclass[twocolumn,english,aps,prb,twocolum,superscriptaddress,bibnotes,amsmath,amssymb,floatfix]{revtex4-2}

\usepackage[colorlinks=true,citecolor=blue,linkcolor=magenta]{hyperref}
\usepackage{changes}
\usepackage{lineno}
\usepackage{stmaryrd}
\usepackage{bm}
\usepackage{graphicx}
\usepackage{epsfig}
\usepackage{color}
\usepackage{amssymb}
\usepackage{amsmath}
\usepackage{array}
\usepackage{braket}
\usepackage{soul}
\usepackage{cleveref}
\usepackage[caption=false]{subfig}
\usepackage[english]{babel}
\usepackage{letltxmacro}
\usepackage{siunitx}
\usepackage{dsfont}

\newcommand{\MPINAT}{Max Planck Institute for Multidisciplinary Sciences, D-37077 G\"{o}ttingen, Germany}
\newcommand{\UGOE}{4th Physical Institute - Solids and Nanostructures, Georg-August-Universit\"{a}t G\"{o}ttingen, D-37077 G\"{o}ttingen, Germany}

\begin{document}

%\preprint{AIP/123-QED}

\title{Coulomb-correlated electron number states in a \\transmission electron microscope beam
}
%\protect\\ i

\author{Rudolf Haindl}
\affiliation{\MPINAT}
\affiliation{\UGOE}
\author{Armin Feist}
\email{armin.feist@mpinat.mpg.de}
\affiliation{\MPINAT}
\affiliation{\UGOE}
\author{Till Domröse}
\affiliation{\MPINAT}
\affiliation{\UGOE}
\author{Marcel Möller}
\affiliation{\MPINAT}
\affiliation{\UGOE}
\author{John H. Gaida}
\affiliation{\MPINAT}
\affiliation{\UGOE}
\author{Sergey\,V. Yalunin}
\affiliation{\MPINAT}
\affiliation{\UGOE}
\author{Claus Ropers}
\email{claus.ropers@mpinat.mpg.de}
\affiliation{\MPINAT}
\affiliation{\UGOE}

\date{\today}

\begin{abstract}
\textbf{We demonstrate the generation of Coulomb-correlated pair, triple and quadruple states of free electrons by femtosecond photoemission from a nanoscale field emitter inside a transmission electron microscope. Event-based electron spectroscopy allows a spatial and spectral characterization of the electron ensemble emitted by each laser pulse. We identify distinctive energy and momentum correlations arising from acceleration-enhanced interparticle energy exchange, revealing strong few-body Coulomb interactions at an energy scale of about two electronvolts. State-sorted beam caustics show a discrete increase in virtual source size and longitudinal source shift for few-electron states, associated with transverse momentum correlations. We observe field-controllable electron antibunching, attributed primarily to transverse Coulomb deflection. The pronounced spatial and spectral characteristics of these electron number states allow filtering schemes that control the statistical distribution of the pulse charge. In this way, the fraction of specific few-electron states can be actively suppressed or enhanced, facilitating the preparation of highly non-Poissonian electron beams for microscopy and lithography, including future heralding schemes and correlated multi-electron probing.}
\end{abstract}

\maketitle

Correlations between electrons are at the core of numerous phenomena in atomic, molecular, and solid-state physics. Mediated by the Coulomb force, few- and many-body electronic correlations govern intriguing phases of matter, such as superconductivity or charge ordering, and they underpin a wide variety of applications, down to nanoscale single-electron sources~\cite{Feve2007, Bocquillon2013} and logic gates based on single charges~\cite{Kastner1992, Zrenner2002}.
In contrast to the opportunities granted by electron correlations in condensed matter, Coulomb interaction in free-electron beams is usually considered a detrimental factor. In electron microscopy, electron repulsion leads to stochastic longitudinal and transverse emittance growth of the beam, described by the Boersch~\cite{Boersch1954} and Loeffler~\cite{Loeffler1969} effects, respectively, and limiting the brightness of state-of-the-art electron sources~\cite{Jansen1990,Cook2010}. In high-charge electron pulses for time-resolved experiments, mean-field and stochastic Coulomb effects govern the achievable pulse duration, energy spread and focusability, and pose a major experimental challenge for ultrafast electron diffraction~\cite{Siwick2002,Collin2005,Michalik2006,Reed2006, Paarmann2012, Ischenko2019} and microscopy~\cite{Cook2016, Feist2017, Bach2019}, particle accelerators~\cite{Hofmann2017} and free-electron lasers~\cite{Emma2006}. 

Studying strong electronic correlations for a beam containing only few particles requires the preparation of a sufficient electron phase space degeneracy. Field emitters represent highly localized sources, and they have been used in studies elucidating free-electron correlations~\cite{Kiesel2002, Kuwahara2021, Keramati2021}. In particular, the physical origin of antibunching in free-electron beams, as reported by Kiesel et al.~\cite{Kiesel2002}, has been a long-standing question and is still actively discussed in the context of exchange-mediated~\cite{Kuwahara2021} and Coulomb~\cite{Baym2014, Kodama2019} interactions. Tailoring such correlations in free-electron beams facilitates sub-Poissonian beam statistics~\cite{Keramati2021}, promising shot-noise reduced imaging and lithography. Strong inter-particle interactions are enabled by spatio-temporally confined femtosecond-pulsed photoemission from nanotips~\cite{Hommelhoff2006,Ropers2007,Barwick2007,Kruger2011, Herink2012, Ciappina2017, Seiffert2018, Dombi2020}, employed for ultrafast electron microscopy and diffraction with high-coherence beams~\cite{Feist2017, Houdellier2018, Paarmann2012}. The pulse-averaged effects of Coulomb interactions from such sources have recently been investigated, associated with spectral broadening and a loss of temporal and spatial resolution~\cite{Yanagisawa2016, Bach2019, Schotz2021, Meier2022a}.

Employing concepts from quantum optics~\cite{Mandel1995}, correlations among free electrons have previously been identified by coincidence detection using detector pairs~\cite{Kiesel2002, Kodama2011, Kuwahara2021, Keramati2021}, as in atomic and molecular science measuring electrons and ions~\cite{Dorner2000, Ullrich2003}, correlated photoemission~\cite{Munoz-Navia2009, vanRiessen2010, Trutzschler2017}, and ionization~\cite{Larochelle1998, Becker2012}.

In electron microscopy, the recent advent of pixelated event detectors has substantially widened the capabilities for coincidence measurements involving electrons, as demonstrated for electron-correlated X-ray emission~\cite{Jannis2019}, cathodoluminescence at nanomaterials~\cite{Varkentina2022} and integrated photonic resonators~\cite{Feist2022}. These capabilities will foster the emerging area of free-electron quantum optics, promising quantum coherent manipulation~\cite{Kfir2021, Asban2021, Zhao2021, Pan2019, BenHayun2021, Ratzel2021} and sensing~\cite{DiGiulio2019, Tsarev2021} at the nanoscale, and facilitating concepts based on electron-electron~\cite{Kiesel2002, Kfir2019, Talebi2021, Konecna2022} or electron-light entanglement~\cite{Kfir2019, Rivera2020, GarciadeAbajo2021, Feist2022, Konecna2022, Taleb2023}. Establishing such schemes will require a fundamental and quantitative understanding of correlations within the single electrons in the beam. 

\begin{figure*}[!ht]
\centering
\includegraphics[width=180mm]{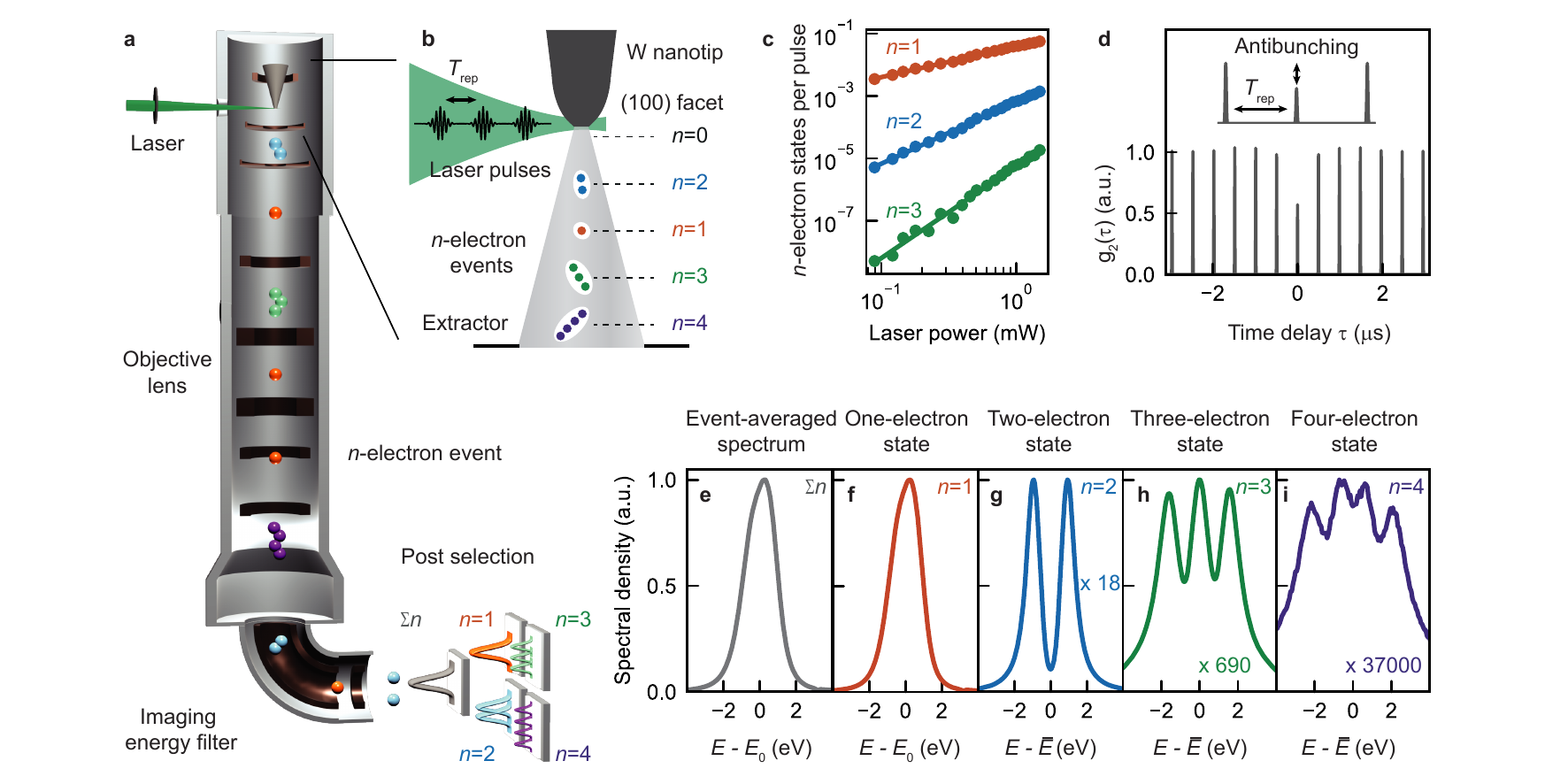}
\caption{\textbf{Electron number states in event-based transmission electron microscopy.} \textbf{a,} Experimental setup. Few-electron states are prepared by pulsed photoemission. The electrons pass the sample plane of the microscope, and post selection in event-based electron spectroscopy enables number state selective beam analysis. \textbf{b,} Ultrashort electron pulses are emitted from a laser-assisted Schottky field emitter (W(110)/ZrO nanotip), with a pulse charge up to few electrons coupled to the microscope column. \textbf{c,} Power-scaling of the rates of one-, two- and three-electron states with fitted slopes of 0.99 (n=1), 1.99 (n=2) and 2.95 (n=3) on a double-logarithmic scale. \textbf{d,} Second-order correlation function $g_2(\tau)$ of detected electrons with a timing resolution of approximately $\SI{10}{\ns}$. The strongly reduced correlation function at zero delay is a clear experimental signature of antibunching. \textbf{e,} The event-averaged spectrum is separated into number-state resolved contributions ($n=1-4$: \textbf{f}-\textbf{i}). The two-, three- and four-electron spectra show a distinct shape with $n$ peaks, indicating a discrete energetic separation.}%
\label{fig_1}
\end{figure*}

Here, we demonstrate strong Coulomb correlations in few-electron states generated at a laser-driven Schottky field emitter.  Using event-based electron spectroscopy and imaging, kinetic energy distributions of electron ensembles emitted by single laser pulses are recorded, sorting events by the number of free electrons. Characteristic multi-lobed spectra for events containing two, three and four electrons are found. We quantitatively characterize inter-particle correlations in both energy and transverse momentum, and observe that these few-body interactions dominate over mean-field (space charge) effects. Two-electron energy correlation functions reveal pronounced peaks separated by around \SI{1.7}{\eV} energy difference, illustrating an energy exchange facilitated by acceleration-enhanced longitudinal interaction along the beam axis. Transverse correlations in conjunction with transverse momentum selection causes antibunching and sub-Poissonian beam statistics. The relative contributions of longitudinal and transverse correlations, and thus the resulting antibunching factor, can be controlled by the initial acceleration field. The findings shed light on fundamental correlations in multi-electron pulses and enable statistical control of electron beams for on-demand correlated few-particle imaging and spectroscopy.

The experiments presented in this study were carried out at the Göttingen Ultrafast Transmission Electron Microscope (see sketch in Fig.~1a)~\cite{Feist2017}. Using a femtosecond laser source, ultrashort electron pulse trains at low pulse charge are generated by near-threshold laser-assisted Schottky emission from a W(100)/ZrO emitter. After propagation through the column of the microscope, the electrons are detected with an event-based camera. The temporal resolution of the electron detector allows for a discrimination of consecutive incident electron pulses, providing an unambiguous measure of the number $n$ of transmitted electrons for each laser pulse (see Fig.~1a,b).

\begin{figure*}[!ht]
\centering
\includegraphics[width=180mm]{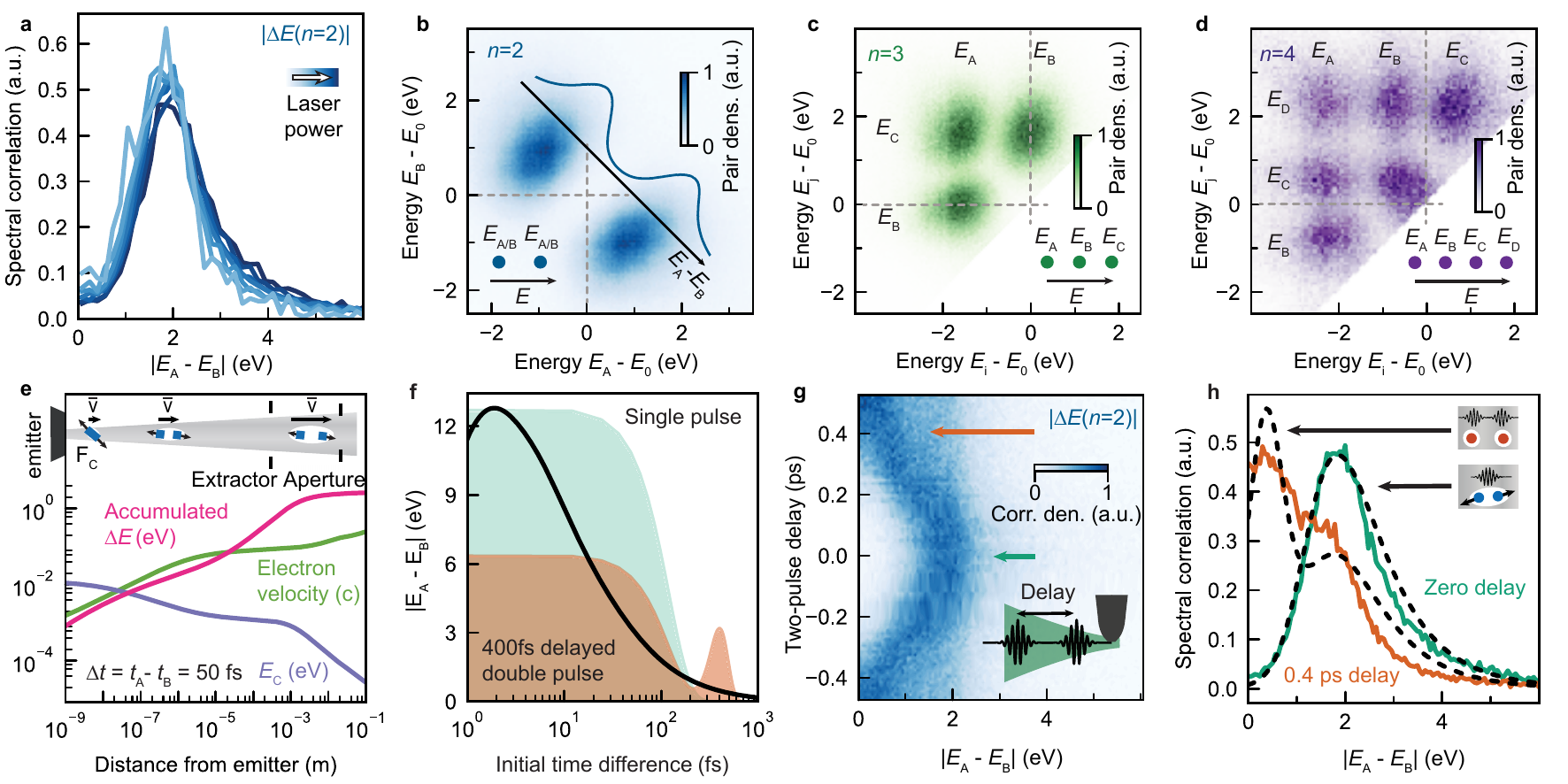}
\caption{\textbf{Coulomb-correlated few-electron pulses.} \textbf{a,} The peak position of normalized one-sided pair correlation functions ($n=2$) is nearly constant for varying laser power. \textbf{b,} Energy histogram of coincident $n=2$-state electron pairs revealing a strong correlation in relative kinetic energy, visible in the spectral correlation function (inset, integrated along the diagonal). \textbf{c$\&$d,} The sorted energy histograms of n=3-states (c) and n=4-states (d) show clearly separated energy-pair correlation peaks of combinations E$_\mathrm{A-C}$ and E$_\mathrm{B-D}$. \textbf{e,} Classical simulation scheme in a geometry consisting of emitter, extraction anode, second acceleration stage and aperture. Two electrons at the nanotip are injected into the static field with a temporal separation of $\Delta t = \SI{50}{fs}$. The momentary Coulomb energy, electron velocity and accumulated energy difference are plotted against electron travel distance from the emitter. A small initial Coulomb energy translates to a greatly enhanced final energy difference during acceleration. \textbf{f,} Plot of final two-electron energy separation for varying emission time difference. Background: Distribution of emission time differences for two Gaussian pulse shapes with a full-width-at-half-maximum width of \SI{200}{\fs} at a delay of \SI{0}{\ps} and \SI{0.4}{\ps}. \textbf{g,} Pair correlation function for photoemission with two delayed laser pulses. At temporal overlap, a strong correlation gap is observed, gradually disappearing for $>$\SI{200}{\fs} pulse delays. \textbf{h,} Comparison of pulse-pair correlation spectra from panel g at temporal overlap and a delay of \SI{0.4}{ps} with simulated correlation spectra (dashed lines).}
\label{fig_2}
\end{figure*}

The rates of $n$-electron events as a function of incident laser power are displayed in Fig.~1c. Specifically, the rate of single-electron emission scales linearly with power, in agreement with the employed process of near-threshold laser-assisted Schottky photoemission~\cite{Cook2009, Yang2010, Feist2017}. Correspondingly, the $n=2$- and $n=3$-electron rates scale with to the power of $n$, i.e., with the square and cube, respectively, of the laser power. (For n=4, only a single measurement at high power was conducted). Considering the relative distribution of $n$-electron events at a given laser power, we identify sub-Poissonian statistics. Specifically, defining $P_1$ as the probability to detect one electron in a pulse, a Poisson process predicts a probability of $P_n = P_1^n /n!$ for detecting $n$ electrons. The actual rates measured for $n=2$ and $n=3$ are lower, at only 85\% and 57\%, respectively, of those expected from the single-electron rate. This antibunching is also evident from a dip in the second-order correlation function at $\tau=0$ (see Fig.~1d), as discussed in detail later in the manuscript.

We next investigate the kinetic energies of these number-sorted electron states (Figs.~1e-h). The spectral distribution of the one-electron events (Fig.~1f), which also dominates the total spectrum (e, summed over all events), consists of a single peak centered around the acceleration energy of $E_0 = \SI{200}{\keV}$. In stark contrast, the spectra of the few-electron events exhibit a number of lobes identical to the number of particles contained. In Figs.~1g-i), we plot the spectral distributions of the electron events sorted into event classes $n=2,3,4$, with respect to the average energy $\overline{E}$ of the electrons in each pulse. Extended Data Fig.~1 displays the spectra of the event classes with respect to the acceleration energy.

For the $n=2$ events, this results in a histogram of energy differences, i.e., the energy correlation function of the two-electron state. Depicted in terms of the magnitude of the energy difference in Fig.~2a, these measurements reveal a clear correlation gap of the energies of both electrons in an $n=2$ state. A natural assumption would be that this energy gap arises from Coulomb interaction. A first question that needs to be answered is to what degree these correlations are modified by emitted electrons near the source which are not transmitted to the column, as such electrons are known to affect the overall spectral distribution~\cite{Kuwahara2016, Jansen1990, Bronsgeest2007, Cook2010, Cook2016, Feist2017, Bach2019}. We find that the correlation function is only weakly dependent on the laser power and thus the average number of electrons (see Extended Data Fig.~2). This shows that we are observing an effect that is governed primarily by the interaction of those few electrons within the measured ensemble.

The measurement scheme further allows us to analyze the spectral characteristics in terms of two-dimensional energy correlation functions. Figure~2b shows the pair-density distribution as a function of the electron energies $E_A$ and $E_B$ associated with two electrons $A$ and $B$ in the same electron pulse. The pair exhibits the strong correlation gap around zero energy difference $E_A-E_B$. The broadening of the pair distribution in the average energy $(E_A+E_B)/2$ is found to depend more strongly on laser power, illustrating that both electrons are affected jointly by an increase of stochastic interactions with electrons not entering the column (see also Methods). An analysis of the pair-distribution of electron energies for the cases of $n=3$ and $n=4$ electrons (Figs.~2c,d) strikingly demonstrates a persistent, regular arrangement of the energies of electrons produced in a single pulse. These measurements further highlight the pronounced interparticle Coulomb correlation at the level of 1-2~eV per electron, which we further study in the following.

In order to elucidate the physical origins of these strong correlations, we numerically simulate the particle propagation including the static acceleration field and inter-particle Coulomb interaction. Specifically, we compute trajectories for sets of electrons with initial conditions representing the emission at the tip, in terms of the distributions of initial momentum, emission location, and temporal separation. Experimental parameters for the acceleration voltages and approximate electrode distances are used in the simulations.

The most important findings of the simulations are a quantitative prediction and rationalization of the magnitude of the observed Coulomb correlations. Figure 2e illustrates the simulation result for an individual pair of electrons emitted with typical parameters for our experimental conditions. The electrons are extracted from the source with a spatial and temporal separation of \SI{8}{\nm} and \SI{50}{\fs}, respectively. At the moment of emission of the second electron, the surface electric field of \SI{0.5}{V/m} has already accelerated the first electron to a distance of \SI{130}{\nm} from the emitter surface, such that the initial transverse separation only accounts for a small fraction of the total particle distance. The electrostatic Coulomb energy at the time of emission of the second electron amounts to only \SI{12}{meV}. Thus, the question arises how such small electrostatic energies can translate to a final energy difference of \SI{2}{eV} and more.

We first provide a qualitative explanation of the enhanced correlation, using non-relativistic expressions for simplicity. In the absence of an accelerating field and for the particles initially at rest, the Coulomb energy $E_{C}$ would only translate into a velocity difference of $\Delta v=v_A-v_B=2\sqrt{E_c/m}$. However, considering the external acceleration of the particles to a mean velocity $\bar{v}=(v_A+v_B)/2$, the same velocity difference results in a kinetic energy difference $\propto 2\bar{v}\Delta v$~\cite{Kruit2009} that is substantially larger than $E_C$. Moreover, Coulomb energy is transferred to high kinetic energy differences only while the electrons are already at higher velocity in the laboratory frame. In particular, for $\bar{v}\gg \Delta v$, the electrons' rate of energy exchange is approximated as the product of the momentary inter-particle Coulomb force and the center-of-mass velocity in the laboratory frame, i.e., $P=F_C \bar{v}$. The final energy difference then becomes $\Delta E=\int dt P(t)$. Therefore, the nearly negligible initial Coulomb energy is magnified by the continuous center-of-mass acceleration to a large final energy difference.

In Figure 2e, the kinetic energy difference (magenta), interparticle Coulomb energy (blue), as well as the momentary electron velocity (green) of the second particle are plotted, on a double-logarithmic scale, as a function of the distance of this particle from the emitter surface. It is evident that a few-hundred meV energy separation emerges upon propagation to the extractor electrode in the electrostatic gun, while a further growth to the final energy difference of nearly \SI{2}{eV} for these particles requires propagation and acceleration over several more millimeters. We note that this scenario represents a maximally controlled version of the Boersch effect in the initial acceleration stages of an electron microscope~\cite{Jansen1988_thesis, Kruit2009}, however, eliminated by the vast majority of its stochastic nature.

\begin{figure}[b!]
\centering
\includegraphics[width=88mm]{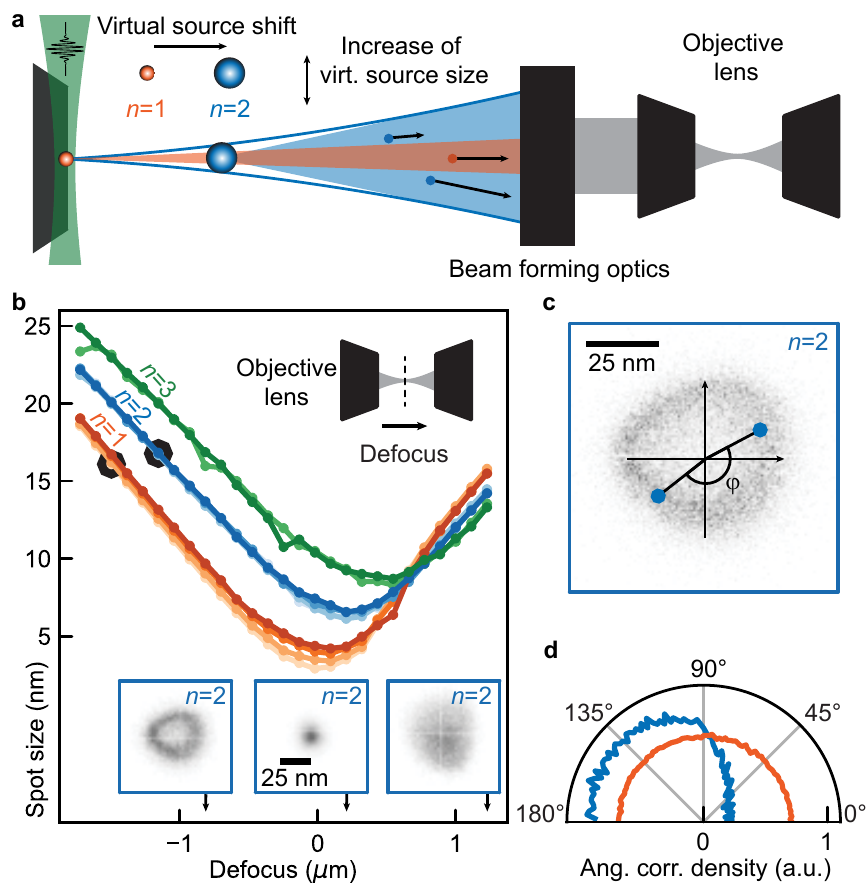}
\caption{\textbf{Characterization of the spatial electron beam properties.} \textbf{a,} Schematic of the effect of Coulomb interaction on an electron beam coupled into an electron microscope. The virtual source increases in size and is shifted along the electron beam axis for pulse charges of two and three electrons. \textbf{b,} Caustics of the electron beam sorted by $n$, recorded by varying the last condenser lens of the microscope (low power: light color, high power: dark color). Insets: images of the beam profile for $n=2$ in underfocus (left), focus (middle) and overfocus (right). \textbf{c,} Image of the beam profile in underfocus, and correlation angle $\varphi$ between electron pairs with respect to the beam center. Long angle legs in the underfocus condition allow for a precise measurement of the angular correlation. \textbf{d,} A strong anisotropic angular correlation is observed for $n=2$, compared to an isotropic distribution for drawing random events from the $n$=1 event class (employed data sets indicated in \textbf{b} by black circles around the data points).}
\label{fig_transverse}
\end{figure}

The simulations also yield further insight into the characteristic timescales over which this electron-electron correlation persists. Figure~2f displays the computed final energy difference as a function of the initial temporal separation of two electrons (black solid line). We find that the energy difference drops to about 1eV within 200 fs. Since the laser pulse acts as a temporal gate for the photoemission, a prediction of the energy correlation function is obtained from these computed energy separations, weighted by the distribution of emission time differences under the photoemission laser pulse envelope (shaded area). The gap then arises from the fact that the laser pulse duration of $\sim150$~fs does not lead to a substantial fraction of electron pairs with a larger separation in emission time. We experimentally probe this interpretation by conducting measurements using a pair of laser pulses of variable delay (see Fig.~2g). The measured correlation gap closes for a temporal separation of the two laser pulses larger than \SI{200}{\fs}. Beyond such delays, an increasing number of electrons with small energy difference and near the central energy of the beam are found, and for those events, one electron is emitted in each pulse. A direct comparison of the experimental and simulated energy correlation functions for pulse overlap and for \SI{400}{\fs} two-pulse delay (Fig.~2g) yields convincing agreement. Furthermore, we find agreement with simulation for n=3,4-states (see Extended Data Fig.~3).

Alongside their spectral distributions and correlations, the few-electron states observed here possess characteristic spatial properties, presented in the following. Specifically, in Fig.~3b, we measure $n$-dependent beam caustics, which exhibit discrete differences in both minimum spot size and focal position. Variations with laser power yield changes to the caustics (higher power leads to some increase in spot size), but are far less pronounced than the differences between the event classes. Under the given conditions, the focusability is limited by spherical aberration of the objective lens and the virtual source size, which result in typical spot profiles for positive and negative defocus (inset in Fig.~3b). Evidently, the $n>1$ caustics are the result of a larger effective source, and the beam waist is shifted towards positive defocus.

Both observations can be understood from mutual transverse deflection (sketch in Fig.~3a), which laterally spreads the few-electron trajectories~\cite{Bach2019}, such that the virtual source increases in size and moves forward, as predicted in simulations~\cite{Kodama2011,Meier2022a}.

A more detailed view of the spatial properties of few-electron states is obtained by analyzing correlations in transverse momentum. To this end, we measure position correlations for a sufficiently large negative defocus (Fig.~3c). The spatial correlation is quantified via the angle $\varphi$ between the two electrons and the beam center. Figure~3d shows the angular correlation density of the two-electron state compared with random correlations drawn from a corresponding single-electron state at the same spot size (\SI{15}{\nm}). In the electron pair state, we obtain a strong anisotropic correlation with a maximum around an angle of $180^{\circ}$, corresponding to electron events localized on opposite sides of the defocused beam, and thus exhibiting nearly opposite transverse momenta. Moreover, the angular correlation becomes most pronounced for events with the largest transverse momentum (see Extended Data Fig.~4).

\begin{figure}[ht!]
\centering
\includegraphics[width=88mm]{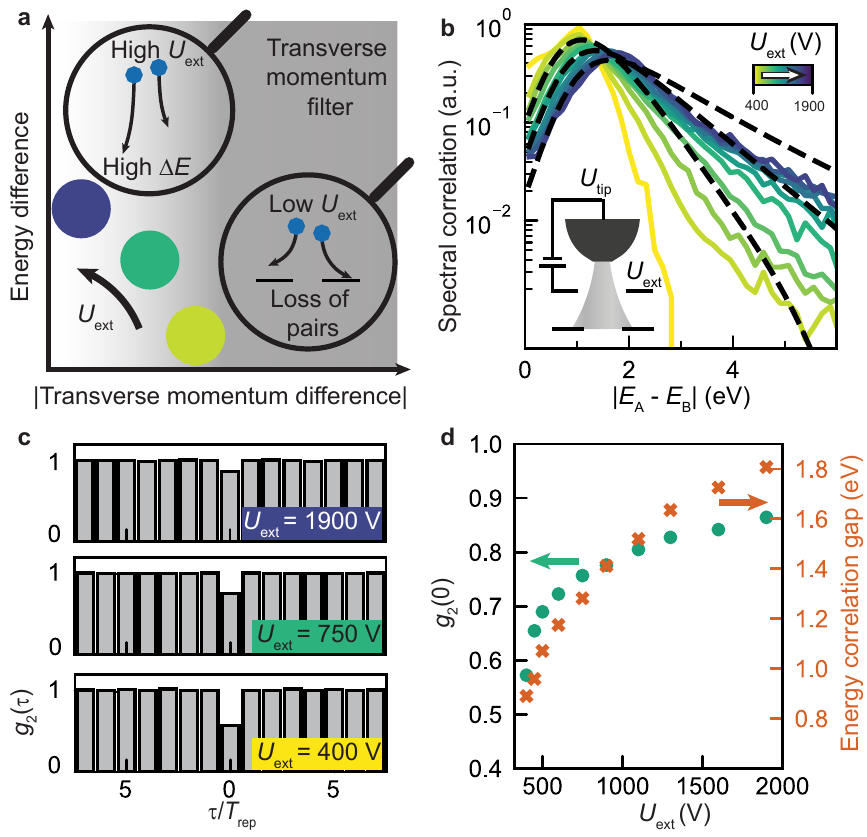}
\caption{\textbf{Electric-field control of longitudinal versus transverse correlations.} \textbf{a,} Via the extraction voltage $U_\mathrm{ext}$, the initial acceleration serves as a control parameter to favor either energy or transverse momentum separation in the doublet states. Electron optical elements act as transverse momentum filters and lead to lower transmission for electron pairs for weaker initial acceleration (low $U_{ext}$). \textbf{b,} Plot of the electron pair correlation functions. Particle tracking simulations (dashed lines) agree with the observed changes in correlation gap with $U_\mathrm{ext}$. \textbf{c,} Delay-dependent current-current correlation function $g_2(\tau)$ for $U_\mathrm{ext}=1900,750,400$~V. The suppression at zero delay ($g_2(0)$) is most pronounced for a low $U_\mathrm{ext}$. \textbf{d,} Plot of $g_2(0)$ (green dots) and energy correlation gap (red dots) vs. extraction voltage.}
\label{fig_antibunching}
\end{figure}

These observations clearly demonstrate that two-particle Coulomb interactions induce pronounced correlations in both the longitudinal and transverse momenta of the electrons. As the correlation primarily emerges in the initial acceleration stages of the electron gun, we explore to what extent they can be controlled by the extraction fields. Qualitatively, a larger acceleration field is expected to enhance the longitudinal correlations and large kinetic energy differences in the beam direction, while a weaker acceleration allows the electrons to exchange more transverse momentum while limiting the growth of the final energy difference. Figure~4a sketches this trade-off between longitudinal and transverse correlations, which manifests itself experimentally in distinct properties of the few-electron states. Specifically, a decrease in extraction voltage, i.e., in the potential difference applied between the tip and the first anode, substantially reduces the observed energy correlation gap and the slope of the high-energy tail (Fig.~4b, solid lines; crosses in Fig.~4d denote the peak of the correlation function). Both features are reproduced in the two-particle simulations described earlier (Fig.~4b, dashed lines).

Interestingly, the enhanced transverse interaction at lower extraction fields has an immediate impact on the statistical distribution of the electron number states. Specifically, the measured beam caustics in Fig.~3b showed that the convergence angle and thus the maximum transverse momentum of the individual particles is the same for all event classes, irrespective of $n$, primarily limited by the microscope's condenser aperture. Therefore, the additional transverse momentum gained by Coulomb repulsion leads to a loss of total transmission of electron pairs. The corresponding change in statistical distribution of events is expressed in terms of the second-order (current-current) correlation function $g_2(\tau)$ as a function of the delay $\tau$ between recorded electrons. Figure 4c shows the second-order current correlation. We see that the emitted charge in sequential pulses is statistically independent ($g_2(\tau)\approx1$), while clear anti-bunching is observed for the electrons recorded from a single pulse ($g_2(0)<1$). In other words, at $U_{ext}=400$~V, the probability of detecting $n=2$ electrons is reduced by a factor of $1-g_2(0)=0.43$ compared to a Poissonian process with the same average electron number per pulse. Importantly, we measure that the antibunching becomes much more pronounced for a smaller extraction voltage (Fig.~4d, circles), illustrating that the enhanced transverse correlations lead to a loss of pairs in the beam path by momentum-selective transmission. As in the case of the energy correlation, the controlled femtosecond temporal gate enabled by photoemission facilitates such strong antibunching~\cite{Keramati2021}, which is orders of magnitude larger than what could be observed for continuous~\cite{Kiesel2002, Kodama2011} and nanosecond-pulsed~\cite{Kuwahara2021} electron microscope beams.

\begin{figure}[ht!]
\centering
\includegraphics[width=88mm]{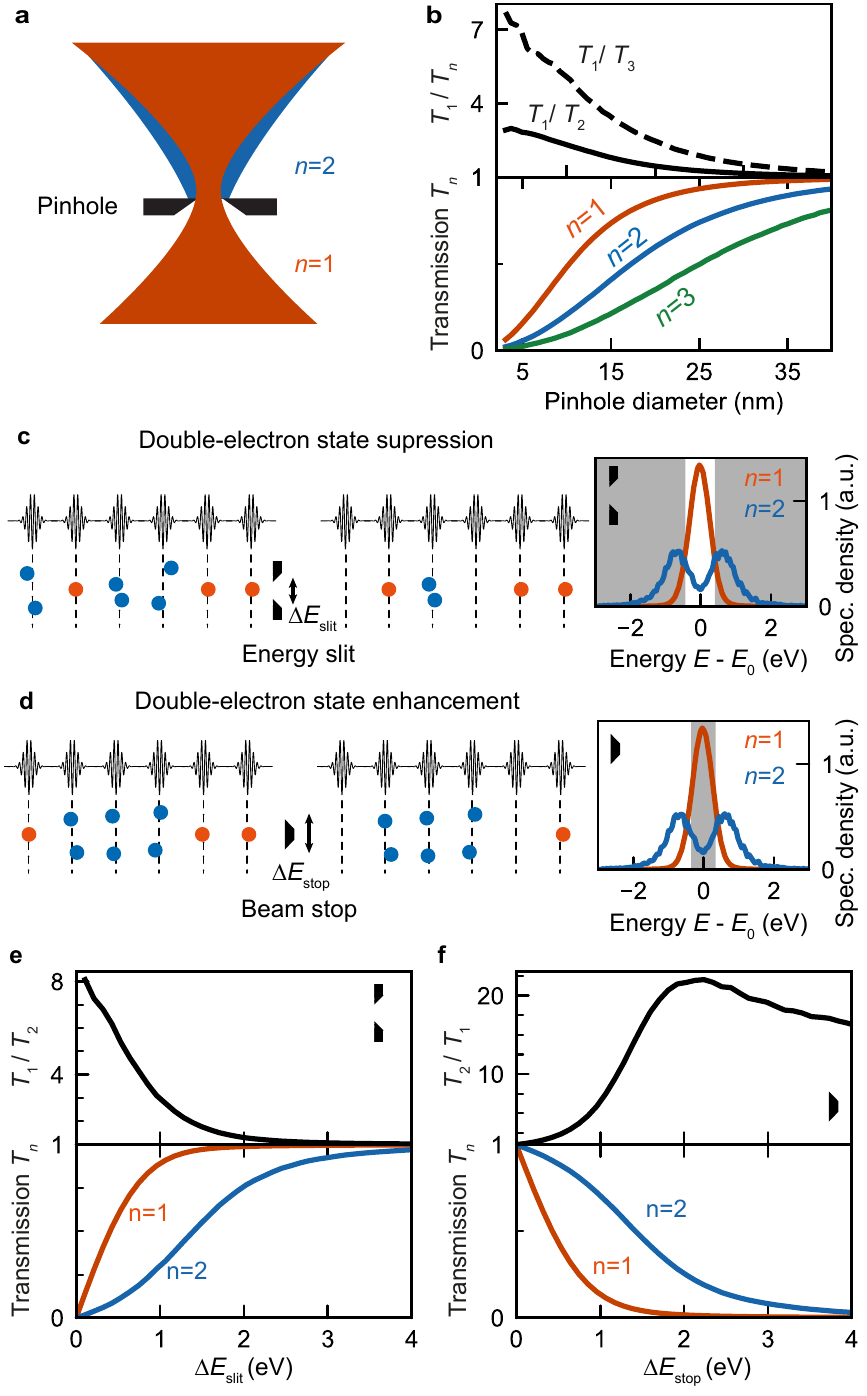}
\caption{\textbf{Statistical control of single- and double-electron states using spatial and spectral filtering.} \textbf{a,} A pinhole positioned at the beam waist of the $n=1$-beam profile spatially filters higher-number states. \textbf{b,} State-selective beam transmissions T$_\mathrm{n}$ (calculated from data in Fig.~\ref{fig_transverse}) and transmission ratios T$_1$ / T$_2$ and T$_1$ / T$_3$, with increased relative selectivity of the $n=1$-electron state. \textbf{c,d,} In a spectrally dispersed plane, an energy-selective slit significantly reduces the transmission of $n=2$-electron states. An energy beam stop suppresses the $n=1$-electron states. Right: Experimental spectra of $n=1$ and $n=2$ electron states (gray area: spectral density rejected by energy slit/beam stop). \textbf{e,f)} Plot of the transmission T$_n$ and transmission ratios T$_1$/T$_2$ and T$_2$/T$_1$, for the scenarios in \textbf{c} and \textbf{d}. Considering individually optimized energy windows, an 8-fold and 20-fold enhanced state-selectivity is found for $n=1$ and $n=2$, respectively.}
\label{fig_5}
\end{figure}

While the employed event-based measurements in conjunction with photoemission gating reveal important aspects of these few-particle correlations, it should be noted that the same phenomena will contribute to the properties of conventional (continuous) electron beams, with direct ramifications for the total beam brightness, coherence, and non-correctable stochastic aberrations. However, in turn, the specific knowledge of these correlations allows for a control of the number statistics in the photoemitted beam, which  may directly benefit microscopy applications. The antibunching observed here and in recent work \cite{Keramati2021} implies that the total photocurrent exhibits sub-Poissonian noise characteristics, a property highly sought after in condensed matter scenarios (e.g. achieved by Coulomb blockade~\cite{Alhassid2000}). In the context of electron microscopy, this feature could be directly applied for shot-noise reduction in imaging, lithography, with immediate consequences for low-dose applications by the possibility to avoid multi-electron specimen damage. In fact, our findings may be directly relevant for the mechanisms underlying the recently observed reduction in sample degradation with pulsed beams~\cite{VandenBussche2019, Kisielowski2019}. Further potential arises from the strong Coulomb-correlations in energy and momentum identified for the few-electron states. For example, the fact that both electrons in the doublet state are well-separated in energy and transverse momentum from each other allows for an energetic or spatial selection of the respective number state. This facilitates a powerful approach to control the statistics of single- and double electron events.

In particular, analyzing the measured spot profiles, a spatial aperture in a beam cross-over could be used to selectively favor the transmission $T_1$ of the $n=1$ number state by a factor of 3 and nearly 8 over the transmissions $T_2$ and $T_3$ of the $n=2$ and $n=3$ states, respectively (Figs.~5a,b). Similarly, a pre-specimen energy filter commonly used in state-of-the-art electron microscopes~\cite{Krivanek2009} could be adjusted to enhance the transmission probability for $n=1$ as compared to $n=2$ states (see Fig.~5c). Specifically, for experimentally measured single-electron and double-electron spectra (Fig.~5e), the $n=1$ transmission probability exceeds the $n=2$ transmission probability by a factor of 8 at small slit widths, greatly amplifying the sub-Poissonian nature of the electron number distribution and facilitating a shot-noise-reduced electron current. 
Conversely, a central beam stop in energy can suppress a substantial fraction of single-electron states, leading to an up to 20-fold enhancement of pair-state over $n=1$ state transmissions (see Fig.~5d,f). This approach will enable new forms of microscopy and spectroscopy with correlated electrons, for a variety of novel two-point or two-time measurement schemes in correlated materials and free-electron quantum optics.

In conclusion, the highly correlated electron number states introduced in this work are of interest both for fundamental considerations and their potential utility in manifold electron beam applications. For example, the pair state can be employed to implement a high-fidelity source of electron-heralded single-electrons, enabling shot-noise-free imaging and lithography with a precisely counted number of electrons. Furthermore, the elementary scattering process creating these well-defined few-body states may generally be assumed to induce entanglement between the electrons. Future studies may address the coherence of such multi-electron states and their possible use as free-electron qubits, with potential applications spanning from interaction-free or correlation-based quantum electron microscopy to quantum information processing.

In the final phase of manuscript preparation, we became aware of a related study by S. Meier, J. Heimerl and P. Hommelhoff~\cite{Meier2022}, who observed energy correlations of photoelectrons emitted from a free-standing tungsten tip.

\clearpage

% redefine figure captions for extended data
\setcounter{figure}{0}
\renewcommand{\figurename}{Extended Data Fig.}
\renewcommand{\thefigure}{\arabic{figure}}

\section*{Methods}

\subsection{Femtosecond electron pulse generation in a transmission electron microscope}
The experimental work was carried out in two commercially available transmission electron microscopes (JEOL JEM 2100F and JEM F200) that have been modified to allow for the investigation of ultrafast dynamics in a stroboscopic laser-pump/electron-probe measurement scheme~\cite{Feist2017}. As our electron source, we employ W/ZrO$_x$(100) Schottky emitters ($r=\SI{490}{\nm}$ radius-of-curvature, $\approx$\SI{100}{\nm} physical emission size) operated at an extraction voltage of $U_\mathrm{ext} =0.4-\SI{2.1}{\kV}$ and a bias voltage of $U_\mathrm{bias} = -\SI{0.3}{\kV}$. After cooling the W(100)/ZrO$_x$ emitter just below the continuous Schottky-emission threshold at \SI{1150}{\K} (filament current \SI{1.6}{\A}), the work function is close to the photon energy of the laser (E$_\mathrm{ph}=$ \SI{2.4}{\eV}, corresponding to \SI{515}{\nm} central wavelength). We generate ultrashort electron pulses via close-to-threshold linear photoemission by focusing laser pulses (\SI{160}{\fs} pulse duration, \SI{600}{\kHz} / \SI{2}{\MHz} repetition rates, $\SI{30}{um} \times \SI{20}{um}$ spot size) onto the apex of the nanotip. While we estimate that every electron pulse initially consists of up to a few hundred electrons~\cite{Bach2019}, apertures in the electro-optical beam path limit the transmitted beam to electrons that were generated close to the optical axis, resulting in average transmitted bunch charges of below one electron per pulse. Subsequent acceleration to \SI{200}{\keV} energy and coupling into the microscope column enable a pulse characterisation in real- and reciprocal space, spectral pulse properties are studied by an imaging energy filter (CEFID, CEOS GmbH).

\subsection{Event-driven photoelectron detection}
The correlated photoelectron states are imaged with a hybrid pixel electron detector that is based on the Timepix3 ASIC (EM CheeTah T3, Amsterdam Scientific Instruments B.V.) and mounted behind the imaging energy filter. The camera generates a stream of data packages containing the position of electron-activated detector pixels, their time-of-arrival (ToA), which are digitized with \SI{1.56}{\ns} time bins, and the energy (time-over-threshold, ToT) associated with incident electron events. At a beam voltage of \SI{200}{\kV} every individual electron activates a cluster of pixels with variable size ($N_\mathrm{pixels, avg} \approx$ 8 pixels), shape and energy ($\mathrm{ToT}_\mathrm{avg} \approx$ 280 $\,$ arb. unit). 

Single-electron event localisation of the ToT-corrected raw data stream is achieved using the Division of Nanoscopy, M4I, Maastricht University event clustering code~\cite{vanSchayck2020,vanSchayckJ.Paul2021}, which is based on a Hierarchical Density-Based Spatial Clustering (HDBSCAN) in Python3. The algorithm reconstructs the timing and position of individual electrons incident on the detector from the activated pixels. Thereby, individual electrons are distinguished in terms of their ToA, attributing between three and nine neighbouring pixels activated within a time window of \SI{100}{\ns} and a summed ToT ranging from 200 $\,$arb. unit to 400 $\,$arb. unit to the same cluster (see Ref.~\cite{vanSchayck2020}).

In a second step, the photoelectrons are clustered according to the femtosecond laser pulse that generated them. The temporal resolution of the detector (\SI{1.56}{\ns}) is much faster than the temporal pulse separation given by the laser ($\SI{1.6}{\us}$), but much slower than the temporal splitting of the correlated electrons at the detector ($\approx \SI{1}{\ps}$). Hereby, the electrons arriving at the detector within $ \Delta t_\mathrm{n} =$\SI{50}{\ns} are assigned to a number-class electron state $n=1,2,3,...$ determined by the number of electrons per laser pulse. The electron correlation time window $\Delta t_\mathrm{n}$ is chosen to capture all correlated electrons while it is much shorter than the dead time between laser pulses.

\subsection{Effect of stochastic Coulomb interactions and mean field on few-electron states.}

Even though only a fraction of electrons generated at the emitter surface is transmitted into the microscope column~\cite{Bach2019}, the spatio-temporal confinement of the emission results in a non-negligible influence of the entire electron cloud on the properties of the transmitted beam. Consequently, mean-field (space charge) as well as stochastic interactions between all electrons need to be considered, and distinguished from the correlations observed in the electron pair state. These different contributions can be assessed in laser-power-dependent measurements. The corresponding $n=1$ and $n=2$ spectral distributions, as well as the $n=2$-average pair energy $(E_A+E_B)/2$ (Extended Data Figs.~2a,b,d) display the expected broadening with increasing laser power (cf. Extended Data Fig.~2e, n=1 broadening: orange circles, average pair energy broadening: grey circles), i.e. scale with the average photocurrent. This is in close correspondence to previous non-event-selective measurements~\cite{Bach2019, Kuwahara2016, Schotz2021} and is typically ascribed to stochastic Coulomb interactions and mean-field effects.

In contrast, the two-electron correlation functions displayed in Extended Data Fig.~2c are remarkably independent of laser power, showing a pronounced gap that is about \SI{1}{\eV} wide, a peak at around \SI{1.8}{\eV}, and an extended tail towards large energy separations exceeding \SI{4}{\eV}. Increasing the photocurrent only imposes moderate variations in the depth of the gap and the shape of the high-energy tail. In particular, the position of the main correlation peak (Extended Data Fig.~2e, blue circles) approaches a fixed value of \SI{1.7}{\eV} towards small average currents, demonstrating that the observed correlation is only weakly altered by multiple Coulomb interactions with the space-charge cloud. Rather, the peak position is dominated by the two-electron correlation.

\subsection{State-average energy subtracted spectra}

Shot-to-shot variations between electron pulses deteriorate the state-averaged (Extended Data Fig.~1a) and number-state resolved (Extended Data Fig.~1b-e) spectra. They are primarily caused by high-voltage and space-charge fluctuations that change the average electron energy $E_0$ for every pulse by an energy $\overline{E}$. As a result, the characteristic multi-peak spectra of the few-electron states is blurred, particularly for electron states with $n \geq 2$. Correcting for $\overline{E}$ for every individual pulse thus significantly enhances the visibility of the multi-peak spectra also for $n=3$ and $n=4$ (cf. Extended Data Fig.~1f-h). The root-mean-square widths of the state-average energies shown in Extended Data Fig.~1 i-k are reduced for higher number states ($n=2,3,4$: $0.73$, $0.6$, $0.52$~eV).

\subsection{Double laser-pulse electron generation}

For the two-laser-pulse generation described in Fig.~2g, a Michelson interferometer splits the incoming laser pulse into two separate pulses. One of the interference arms has a variable optical path length, implemented by a retroreflector mounted on a delay stage. The delay time of the two optical pulses (up to \SI{2}{\ps}) is much lower than the laser pulse repetition time ($\SI{1.6}{us}$, corresponding to a repetition rate of \SI{600}{\kHz}). Hence, two photoelectrons generated by two separate laser pulses and two photoelectrons generated by the same pulse are both detected as two-electron events.

As the optical power on the tip oscillates for small delay times due to constructive and destructive interference of the laser pulses, the number of generated electrons strongly varies in this delay regime. Therefore, we selected delays with approximately the same one-electron-state rate ($\pm \sigma / 2$) over the integration time of five seconds.\\

\subsection{Numerical simulation of multi-particle trajectories}

Energy correlation histograms for the electron number states $n=2-4$ are shown in Fig.~2b-d.  These correlation spectra can be reproduced with the numerical multi-particle trajectory simulations discussed in Fig.~2e-h. For the simulation of the $n=3,4$-correlation spectra, the model is extended to three and four particles. We compute the electron trajectories of all $n$-states for a set of parameters within the experimental range, i.e., an extraction voltage of \SI{2100}{V}, a temporal emission profile of \SI{180}{\fs}, a physical source size of \SI{100}{\nm} and considering the mean-field broadening of \SI{1}{\eV} observed for the $n=1$ state. The simulated multi-particle energy-pair histograms are shown in Fig.~3(d-f) and are in excellent agreement with the experimental data (Fig.~3a-c) in terms of the observed correlation gaps and peak positions.

\noindent\textbf{Acknowledgements}
We gratefully acknowledge P. Kruit for fruitful discussions on stochastic Coulomb interactions. We are indebted to the members of the G\"{o}ttingen UTEM team for constant support and useful discussions. 

\noindent\textbf{Data availability statement}
The code and data used to produce the plots within this work will be released on the repository Edmond upon publication of this preprint.

\noindent\textbf{Funding Information:}
The work at the Göttingen UTEM Lab was funded by the Deutsche Forschungsgemeinschaft (DFG, German Research Foundation) through  432680300/SFB\,1456 (project C01) and the Gottfried Wilhelm Leibniz program.

\noindent\textbf{Author contribution:}
Conceptualization: C.R., and A.F. Methodology: R.H., and A.F. (event-based spectroscopy), R.H., J.H.G., M.M., and A.F. (photoemission gun), S.V.Y., R.H., T.D., A.F., and C.R. (particle simulation). Investigation: R.H., A.F., M.M. (experiment), S.V.Y., R.H., C.R., T.D. (simulations). Formal Analysis: R.H., A.F., and C.R. Data curation: R.H. Software: R.H., M.M., A.F., and J.H.G. Project administration, supervision, and funding acquisition: C.R. Visualization: R.H. and T.D Writing: R.H., A.F., S.V.Y, and C.R., with comments and feedback from all authors. \\

\onecolumngrid

\begin{figure*}[!ht]
\centering
\includegraphics[width=88mm]{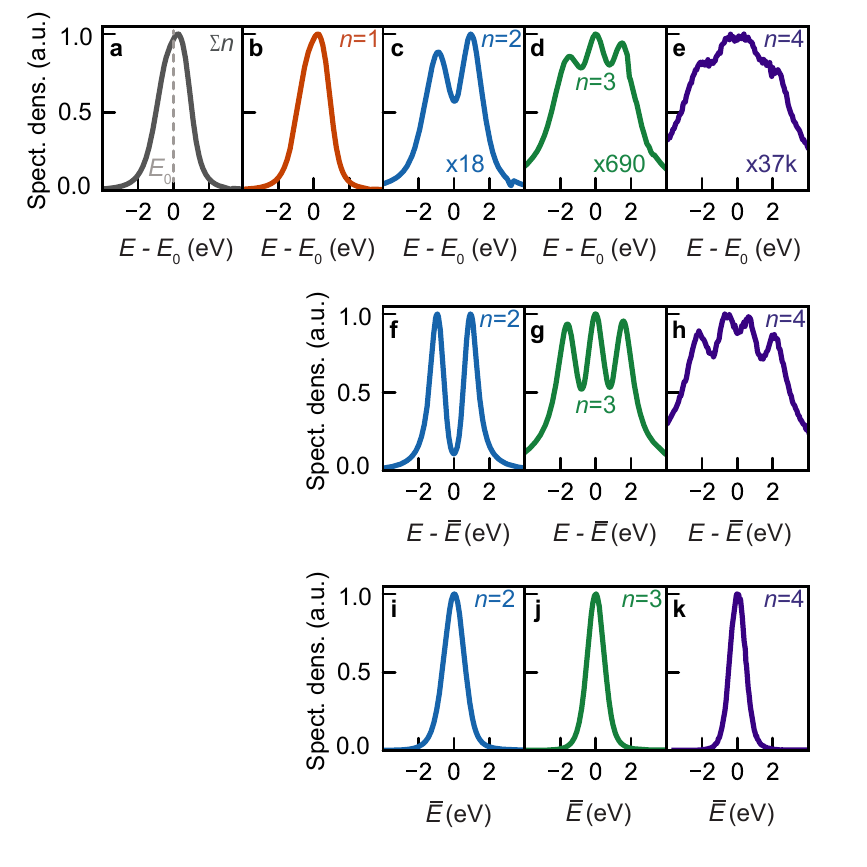}
\caption{\textbf{State-average energy subtracted spectra.} The event-averaged spectrum with a rms energy spread of \SI{1.08}{eV} (\textbf{a}) is separated into number-state resolved contributions ($n=1-4$: \textbf{b}-\textbf{e}). Voltage- and space charge-fluctuations smear out the characteristic n-peak spectra. \textbf{f-h,} For $n=2-4$, the state-average energy $\overline{E}$ is subtracted from the $n$-state energies and the $\overline{E}$-corrected $n$-spectra are plotted. \textbf{i-k,} Plot of the state-average spectra for $n=2-4$ with a rms energy spread of $n=2$: \SI{0.73}{eV}, $n=3$: \SI{0.60}{eV}, $n=3$: \SI{0.52}{eV}.}%
\label{fig_dejittering}
\end{figure*}

\clearpage

\begin{figure*}[!ht]
\centering
\includegraphics[width=88mm]{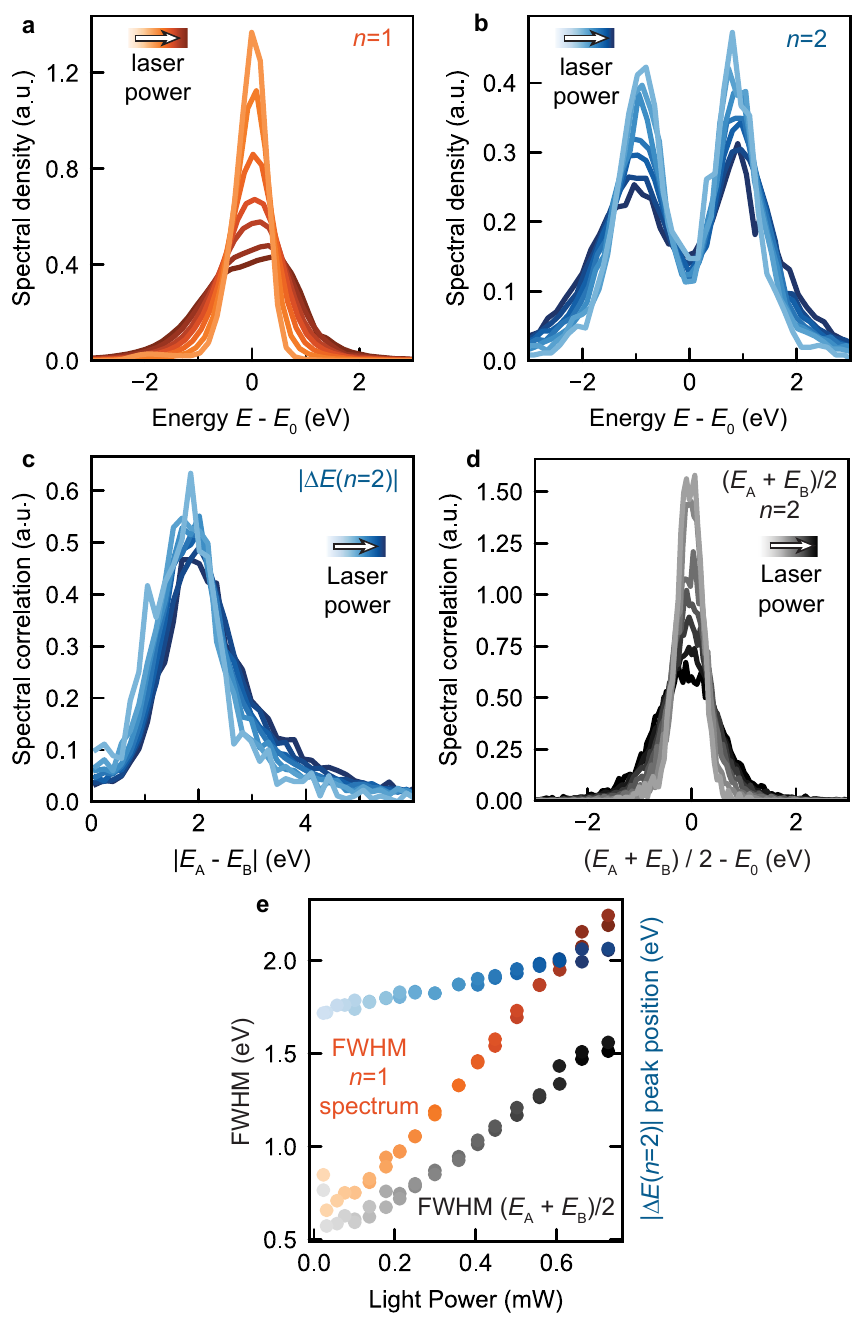}
\caption{\textbf{Effect of stochastic Coulomb interactions and space charge on few-electron states. a, b,} Normalized $n=1$-spectra (a) and $n=2$-spectra (b) for varying laser power. \textbf{c,d,} Normalized one-sided pair correlation functions (c) and pair distributions in average energy $(E_A+E_B)/2$ (d) (both for $n=2$) for varying laser power. \textbf{e,} Power scaling of the peak position of the n = 2-correlation function compared to the spectral width (FWHM) of the $n=1$-state and of the electron pair average energy.}%
\label{fig_space_charge}
\end{figure*}

\clearpage

\begin{figure*}
\centering
\includegraphics[width=180mm]{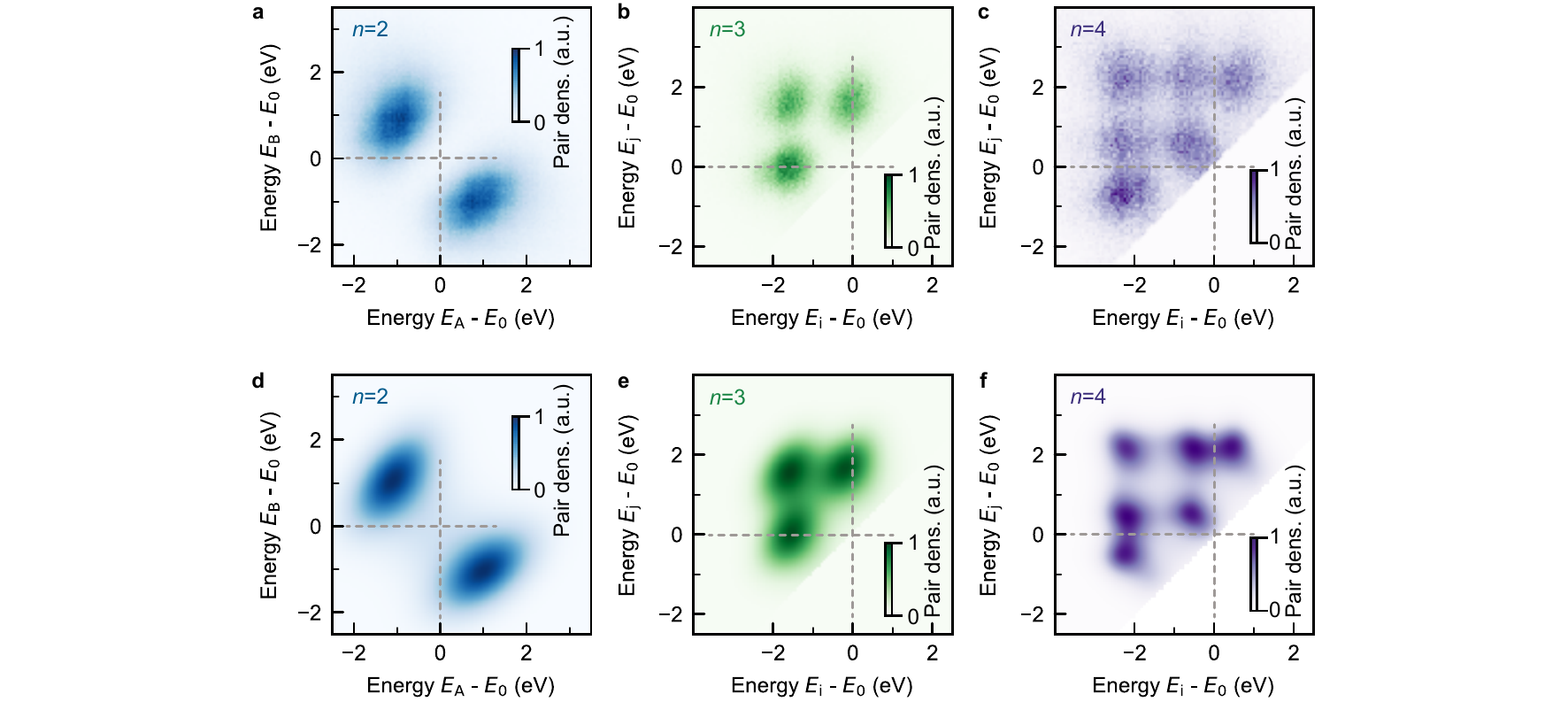}
\caption{\textbf{Numerical simulations for double-, triple- and quadruple-states.} The experimental sorted energy histograms of the $n=2-4$-states (a-c) are compared with multi-electron energy histograms from the particle trajectory simulation (d-f: $n=2-4$). The energy-pair correlation peaks are clearly resolved and in excellent agreement with the experimental data.}%
\label{fig_multi_electron_simulation}
\end{figure*}

\clearpage

\begin{figure*}
\centering
\includegraphics[width=180mm]{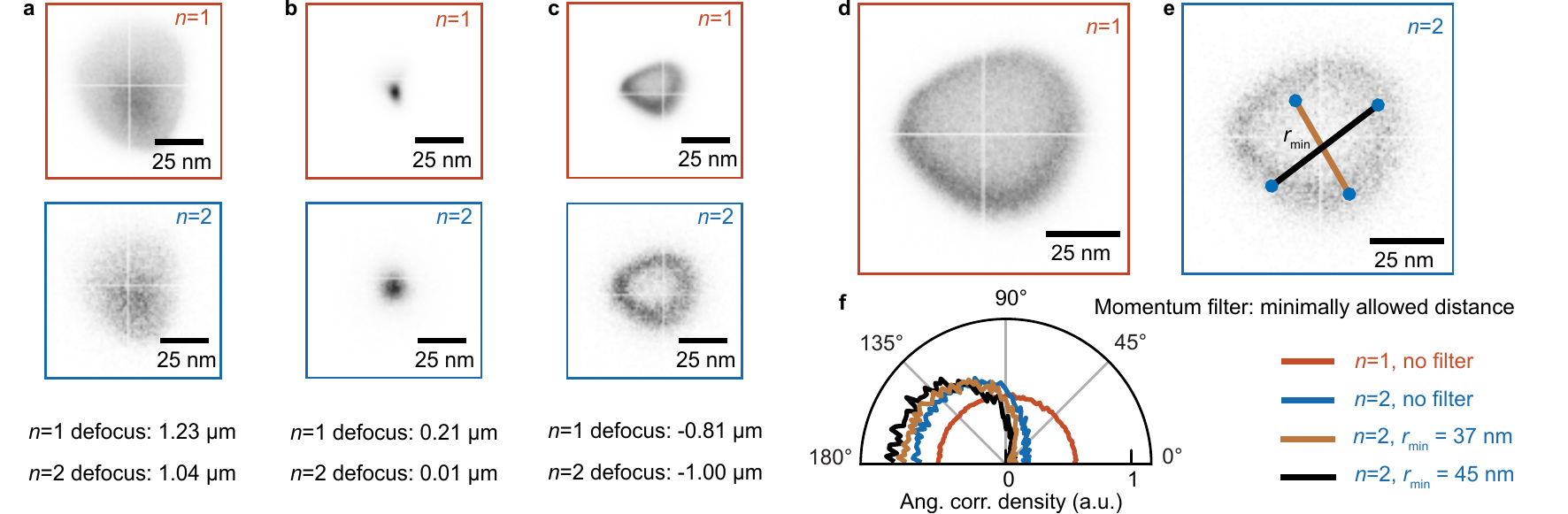}
\caption{\textbf{Comparison of beam profiles and transverse momentum filter. a-c,} Beam profiles of $n=$1 and $n=$2 states in overfocus (a), focus (b) and underfocus (c) condition. The beam profiles of both number classes agree qualitatively considering the  Coulomb-correlation induced defocus shift and increase in spot size. \textbf{d, e} Underfocus $n=1$ and $n=2$ beam profiles used in the angular correlation analysis in Fig.~4d in the main text. \textbf{e,} Digital transverse momentum filters are plotted in brown and black onto the $n=2$-beam profile. The filters only considers electron pairs where both electrons have a minimum distance $r_\mathrm{min}$ to each other. \textbf{f,} Plot of the angular correlation density without filter for $n=1,2$ (red and blue) and with filter for $n=2$ (brown: $r_\mathrm{min}=\SI{37}{nm}$ and  black: $r_\mathrm{min}=\SI{45}{nm}$). The distribution for $n=2$ becomes increasingly  more pronounced at \SI{180}{\degree} after application of the transverse momentum filters.}%
\label{fig_transverse_extended_data}
\end{figure*}

\clearpage

\twocolumngrid

\bibliography{arxiv_submission_update_april2023}% Produces the bibliography via BibTeX.

\begin{thebibliography}{75}%
\makeatletter
\providecommand \@ifxundefined [1]{%
 \@ifx{#1\undefined}
}%
\providecommand \@ifnum [1]{%
 \ifnum #1\expandafter \@firstoftwo
 \else \expandafter \@secondoftwo
 \fi
}%
\providecommand \@ifx [1]{%
 \ifx #1\expandafter \@firstoftwo
 \else \expandafter \@secondoftwo
 \fi
}%
\providecommand \natexlab [1]{#1}%
\providecommand \enquote  [1]{``#1''}%
\providecommand \bibnamefont  [1]{#1}%
\providecommand \bibfnamefont [1]{#1}%
\providecommand \citenamefont [1]{#1}%
\providecommand \href@noop [0]{\@secondoftwo}%
\providecommand \href [0]{\begingroup \@sanitize@url \@href}%
\providecommand \@href[1]{\@@startlink{#1}\@@href}%
\providecommand \@@href[1]{\endgroup#1\@@endlink}%
\providecommand \@sanitize@url [0]{\catcode `\\12\catcode `\$12\catcode
  `\&12\catcode `\#12\catcode `\^12\catcode `\_12\catcode `\%12\relax}%
\providecommand \@@startlink[1]{}%
\providecommand \@@endlink[0]{}%
\providecommand \url  [0]{\begingroup\@sanitize@url \@url }%
\providecommand \@url [1]{\endgroup\@href {#1}{\urlprefix }}%
\providecommand \urlprefix  [0]{URL }%
\providecommand \Eprint [0]{\href }%
\providecommand \doibase [0]{https://doi.org/}%
\providecommand \selectlanguage [0]{\@gobble}%
\providecommand \bibinfo  [0]{\@secondoftwo}%
\providecommand \bibfield  [0]{\@secondoftwo}%
\providecommand \translation [1]{[#1]}%
\providecommand \BibitemOpen [0]{}%
\providecommand \bibitemStop [0]{}%
\providecommand \bibitemNoStop [0]{.\EOS\space}%
\providecommand \EOS [0]{\spacefactor3000\relax}%
\providecommand \BibitemShut  [1]{\csname bibitem#1\endcsname}%
\let\auto@bib@innerbib\@empty
%</preamble>
\bibitem [{\citenamefont {F{\`e}ve}\ \emph {et~al.}(2007)\citenamefont
  {F{\`e}ve}, \citenamefont {Mah{\'e}}, \citenamefont {Berroir}, \citenamefont
  {Kontos}, \citenamefont {Pla{\c c}ais}, \citenamefont {Glattli},
  \citenamefont {Cavanna}, \citenamefont {Etienne},\ and\ \citenamefont
  {Jin}}]{Feve2007}%
  \BibitemOpen
  \bibfield  {author} {\bibinfo {author} {\bibfnamefont {G.}~\bibnamefont
  {F{\`e}ve}}, \bibinfo {author} {\bibfnamefont {A.}~\bibnamefont {Mah{\'e}}},
  \bibinfo {author} {\bibfnamefont {J.-M.}\ \bibnamefont {Berroir}}, \bibinfo
  {author} {\bibfnamefont {T.}~\bibnamefont {Kontos}}, \bibinfo {author}
  {\bibfnamefont {B.}~\bibnamefont {Pla{\c c}ais}}, \bibinfo {author}
  {\bibfnamefont {D.~C.}\ \bibnamefont {Glattli}}, \bibinfo {author}
  {\bibfnamefont {A.}~\bibnamefont {Cavanna}}, \bibinfo {author} {\bibfnamefont
  {B.}~\bibnamefont {Etienne}},\ and\ \bibinfo {author} {\bibfnamefont
  {Y.}~\bibnamefont {Jin}},\ }\href {https://doi.org/10.1126/science.1141243}
  {\bibfield  {journal} {\bibinfo  {journal} {Science}\ }\textbf {\bibinfo
  {volume} {316}},\ \bibinfo {pages} {1169} (\bibinfo {year}
  {2007})}\BibitemShut {NoStop}%
\bibitem [{\citenamefont {Bocquillon}\ \emph {et~al.}(2013)\citenamefont
  {Bocquillon}, \citenamefont {Freulon}, \citenamefont {Berroir}, \citenamefont
  {Degiovanni}, \citenamefont {Pla{\c c}ais}, \citenamefont {Cavanna},
  \citenamefont {Jin},\ and\ \citenamefont {F{\`e}ve}}]{Bocquillon2013}%
  \BibitemOpen
  \bibfield  {author} {\bibinfo {author} {\bibfnamefont {E.}~\bibnamefont
  {Bocquillon}}, \bibinfo {author} {\bibfnamefont {V.}~\bibnamefont {Freulon}},
  \bibinfo {author} {\bibfnamefont {J.-M.}\ \bibnamefont {Berroir}}, \bibinfo
  {author} {\bibfnamefont {P.}~\bibnamefont {Degiovanni}}, \bibinfo {author}
  {\bibfnamefont {B.}~\bibnamefont {Pla{\c c}ais}}, \bibinfo {author}
  {\bibfnamefont {A.}~\bibnamefont {Cavanna}}, \bibinfo {author} {\bibfnamefont
  {Y.}~\bibnamefont {Jin}},\ and\ \bibinfo {author} {\bibfnamefont
  {G.}~\bibnamefont {F{\`e}ve}},\ }\href
  {https://doi.org/10.1126/science.1232572} {\bibfield  {journal} {\bibinfo
  {journal} {Science}\ }\textbf {\bibinfo {volume} {339}},\ \bibinfo {pages}
  {1054} (\bibinfo {year} {2013})}\BibitemShut {NoStop}%
\bibitem [{\citenamefont {Kastner}(1992)}]{Kastner1992}%
  \BibitemOpen
  \bibfield  {author} {\bibinfo {author} {\bibfnamefont {M.~A.}\ \bibnamefont
  {Kastner}},\ }\href {https://doi.org/10.1103/RevModPhys.64.849} {\bibfield
  {journal} {\bibinfo  {journal} {Reviews of Modern Physics}\ }\textbf
  {\bibinfo {volume} {64}},\ \bibinfo {pages} {849} (\bibinfo {year}
  {1992})}\BibitemShut {NoStop}%
\bibitem [{\citenamefont {Zrenner}\ \emph {et~al.}(2002)\citenamefont
  {Zrenner}, \citenamefont {Beham}, \citenamefont {Stufler}, \citenamefont
  {Findeis}, \citenamefont {Bichler},\ and\ \citenamefont
  {Abstreiter}}]{Zrenner2002}%
  \BibitemOpen
  \bibfield  {author} {\bibinfo {author} {\bibfnamefont {A.}~\bibnamefont
  {Zrenner}}, \bibinfo {author} {\bibfnamefont {E.}~\bibnamefont {Beham}},
  \bibinfo {author} {\bibfnamefont {S.}~\bibnamefont {Stufler}}, \bibinfo
  {author} {\bibfnamefont {F.}~\bibnamefont {Findeis}}, \bibinfo {author}
  {\bibfnamefont {M.}~\bibnamefont {Bichler}},\ and\ \bibinfo {author}
  {\bibfnamefont {G.}~\bibnamefont {Abstreiter}},\ }\href
  {https://doi.org/10.1038/nature00912} {\bibfield  {journal} {\bibinfo
  {journal} {Nature}\ }\textbf {\bibinfo {volume} {418}},\ \bibinfo {pages}
  {612} (\bibinfo {year} {2002})}\BibitemShut {NoStop}%
\bibitem [{\citenamefont {Boersch}(1954)}]{Boersch1954}%
  \BibitemOpen
  \bibfield  {author} {\bibinfo {author} {\bibfnamefont {H.}~\bibnamefont
  {Boersch}},\ }\href {https://doi.org/10.1007/BF01375256} {\bibfield
  {journal} {\bibinfo  {journal} {Zeitschrift f\"ur Physik}\ }\textbf {\bibinfo
  {volume} {139}},\ \bibinfo {pages} {115} (\bibinfo {year}
  {1954})}\BibitemShut {NoStop}%
\bibitem [{\citenamefont {Loeffler}(1969)}]{Loeffler1969}%
  \BibitemOpen
  \bibfield  {author} {\bibinfo {author} {\bibfnamefont {{\relax
  KH}.}~\bibnamefont {Loeffler}},\ }\href@noop {} {\bibfield  {journal}
  {\bibinfo  {journal} {ZEITSCHRIFT FUR ANGEWANDTE PHYSIK}\ }\textbf {\bibinfo
  {volume} {27}},\ \bibinfo {pages} {145} (\bibinfo {year} {1969})}\BibitemShut
  {NoStop}%
\bibitem [{\citenamefont {Jansen}(1990)}]{Jansen1990}%
  \BibitemOpen
  \bibfield  {author} {\bibinfo {author} {\bibfnamefont {G.}~\bibnamefont
  {Jansen}},\ }\href {https://doi.org/10.1016/0168-9002(90)90652-M} {\bibfield
  {journal} {\bibinfo  {journal} {Nuclear Instruments and Methods in Physics
  Research Section A: Accelerators, Spectrometers, Detectors and Associated
  Equipment}\ }\textbf {\bibinfo {volume} {298}},\ \bibinfo {pages} {496}
  (\bibinfo {year} {1990})}\BibitemShut {NoStop}%
\bibitem [{\citenamefont {Cook}\ \emph {et~al.}(2010)\citenamefont {Cook},
  \citenamefont {Verduin}, \citenamefont {Hagen},\ and\ \citenamefont
  {Kruit}}]{Cook2010}%
  \BibitemOpen
  \bibfield  {author} {\bibinfo {author} {\bibfnamefont {B.}~\bibnamefont
  {Cook}}, \bibinfo {author} {\bibfnamefont {T.}~\bibnamefont {Verduin}},
  \bibinfo {author} {\bibfnamefont {C.~W.}\ \bibnamefont {Hagen}},\ and\
  \bibinfo {author} {\bibfnamefont {P.}~\bibnamefont {Kruit}},\ }\href
  {https://doi.org/10.1116/1.3502642} {\bibfield  {journal} {\bibinfo
  {journal} {Journal of Vacuum Science \& Technology B, Nanotechnology and
  Microelectronics: Materials, Processing, Measurement, and Phenomena}\
  }\textbf {\bibinfo {volume} {28}},\ \bibinfo {pages} {C6C74} (\bibinfo {year}
  {2010})}\BibitemShut {NoStop}%
\bibitem [{\citenamefont {Siwick}\ \emph {et~al.}(2002)\citenamefont {Siwick},
  \citenamefont {Dwyer}, \citenamefont {Jordan},\ and\ \citenamefont
  {Miller}}]{Siwick2002}%
  \BibitemOpen
  \bibfield  {author} {\bibinfo {author} {\bibfnamefont {B.~J.}\ \bibnamefont
  {Siwick}}, \bibinfo {author} {\bibfnamefont {J.~R.}\ \bibnamefont {Dwyer}},
  \bibinfo {author} {\bibfnamefont {R.~E.}\ \bibnamefont {Jordan}},\ and\
  \bibinfo {author} {\bibfnamefont {R.~J.~D.}\ \bibnamefont {Miller}},\ }\href
  {https://doi.org/10.1063/1.1487437} {\bibfield  {journal} {\bibinfo
  {journal} {Journal of Applied Physics}\ }\textbf {\bibinfo {volume} {92}},\
  \bibinfo {pages} {1643} (\bibinfo {year} {2002})}\BibitemShut {NoStop}%
\bibitem [{\citenamefont {Collin}\ \emph {et~al.}(2005)\citenamefont {Collin},
  \citenamefont {Merano}, \citenamefont {Gatri}, \citenamefont {Sonderegger},
  \citenamefont {Renucci}, \citenamefont {Gani{\`e}re},\ and\ \citenamefont
  {Deveaud}}]{Collin2005}%
  \BibitemOpen
  \bibfield  {author} {\bibinfo {author} {\bibfnamefont {S.}~\bibnamefont
  {Collin}}, \bibinfo {author} {\bibfnamefont {M.}~\bibnamefont {Merano}},
  \bibinfo {author} {\bibfnamefont {M.}~\bibnamefont {Gatri}}, \bibinfo
  {author} {\bibfnamefont {S.}~\bibnamefont {Sonderegger}}, \bibinfo {author}
  {\bibfnamefont {P.}~\bibnamefont {Renucci}}, \bibinfo {author} {\bibfnamefont
  {J.-D.}\ \bibnamefont {Gani{\`e}re}},\ and\ \bibinfo {author} {\bibfnamefont
  {B.}~\bibnamefont {Deveaud}},\ }\href {https://doi.org/10.1063/1.2128494}
  {\bibfield  {journal} {\bibinfo  {journal} {Journal of Applied Physics}\
  }\textbf {\bibinfo {volume} {98}},\ \bibinfo {pages} {094910} (\bibinfo
  {year} {2005})}\BibitemShut {NoStop}%
\bibitem [{\citenamefont {Michalik}\ and\ \citenamefont
  {Sipe}(2006)}]{Michalik2006}%
  \BibitemOpen
  \bibfield  {author} {\bibinfo {author} {\bibfnamefont {A.~M.}\ \bibnamefont
  {Michalik}}\ and\ \bibinfo {author} {\bibfnamefont {J.~E.}\ \bibnamefont
  {Sipe}},\ }\href {https://doi.org/10.1063/1.2178855} {\bibfield  {journal}
  {\bibinfo  {journal} {Journal of Applied Physics}\ }\textbf {\bibinfo
  {volume} {99}},\ \bibinfo {pages} {054908} (\bibinfo {year}
  {2006})}\BibitemShut {NoStop}%
\bibitem [{\citenamefont {Reed}(2006)}]{Reed2006}%
  \BibitemOpen
  \bibfield  {author} {\bibinfo {author} {\bibfnamefont {B.~W.}\ \bibnamefont
  {Reed}},\ }\href {https://doi.org/10.1063/1.2227710} {\bibfield  {journal}
  {\bibinfo  {journal} {Journal of Applied Physics}\ }\textbf {\bibinfo
  {volume} {100}},\ \bibinfo {pages} {034916} (\bibinfo {year}
  {2006})}\BibitemShut {NoStop}%
\bibitem [{\citenamefont {Paarmann}\ \emph {et~al.}(2012)\citenamefont
  {Paarmann}, \citenamefont {Gulde}, \citenamefont {M{\"u}ller}, \citenamefont
  {Sch{\"a}fer}, \citenamefont {Schweda}, \citenamefont {Maiti}, \citenamefont
  {Xu}, \citenamefont {Hohage}, \citenamefont {Schenk}, \citenamefont
  {Ropers},\ and\ \citenamefont {Ernstorfer}}]{Paarmann2012}%
  \BibitemOpen
  \bibfield  {author} {\bibinfo {author} {\bibfnamefont {A.}~\bibnamefont
  {Paarmann}}, \bibinfo {author} {\bibfnamefont {M.}~\bibnamefont {Gulde}},
  \bibinfo {author} {\bibfnamefont {M.}~\bibnamefont {M{\"u}ller}}, \bibinfo
  {author} {\bibfnamefont {S.}~\bibnamefont {Sch{\"a}fer}}, \bibinfo {author}
  {\bibfnamefont {S.}~\bibnamefont {Schweda}}, \bibinfo {author} {\bibfnamefont
  {M.}~\bibnamefont {Maiti}}, \bibinfo {author} {\bibfnamefont
  {C.}~\bibnamefont {Xu}}, \bibinfo {author} {\bibfnamefont {T.}~\bibnamefont
  {Hohage}}, \bibinfo {author} {\bibfnamefont {F.}~\bibnamefont {Schenk}},
  \bibinfo {author} {\bibfnamefont {C.}~\bibnamefont {Ropers}},\ and\ \bibinfo
  {author} {\bibfnamefont {R.}~\bibnamefont {Ernstorfer}},\ }\href
  {https://doi.org/10.1063/1.4768204} {\bibfield  {journal} {\bibinfo
  {journal} {Journal of Applied Physics}\ }\textbf {\bibinfo {volume} {112}},\
  \bibinfo {pages} {113109} (\bibinfo {year} {2012})}\BibitemShut {NoStop}%
\bibitem [{\citenamefont {Ischenko}\ \emph {et~al.}(2019)\citenamefont
  {Ischenko}, \citenamefont {Kochikov},\ and\ \citenamefont
  {Miller}}]{Ischenko2019}%
  \BibitemOpen
  \bibfield  {author} {\bibinfo {author} {\bibfnamefont {A.~A.}\ \bibnamefont
  {Ischenko}}, \bibinfo {author} {\bibfnamefont {I.~V.}\ \bibnamefont
  {Kochikov}},\ and\ \bibinfo {author} {\bibfnamefont {R.~J.~D.}\ \bibnamefont
  {Miller}},\ }\href {https://doi.org/10.1063/1.5060673} {\bibfield  {journal}
  {\bibinfo  {journal} {The Journal of Chemical Physics}\ }\textbf {\bibinfo
  {volume} {150}},\ \bibinfo {pages} {054201} (\bibinfo {year}
  {2019})}\BibitemShut {NoStop}%
\bibitem [{\citenamefont {Cook}\ and\ \citenamefont {Kruit}(2016)}]{Cook2016}%
  \BibitemOpen
  \bibfield  {author} {\bibinfo {author} {\bibfnamefont {B.}~\bibnamefont
  {Cook}}\ and\ \bibinfo {author} {\bibfnamefont {P.}~\bibnamefont {Kruit}},\
  }\href {https://doi.org/10.1063/1.4963783} {\bibfield  {journal} {\bibinfo
  {journal} {Applied Physics Letters}\ }\textbf {\bibinfo {volume} {109}},\
  \bibinfo {pages} {151901} (\bibinfo {year} {2016})}\BibitemShut {NoStop}%
\bibitem [{\citenamefont {Feist}\ \emph {et~al.}(2017)\citenamefont {Feist},
  \citenamefont {Bach}, \citenamefont {{Rubiano da Silva}}, \citenamefont
  {Danz}, \citenamefont {M{\"o}ller}, \citenamefont {Priebe}, \citenamefont
  {Domr{\"o}se}, \citenamefont {Gatzmann}, \citenamefont {Rost}, \citenamefont
  {Schauss}, \citenamefont {Strauch}, \citenamefont {Bormann}, \citenamefont
  {Sivis}, \citenamefont {Sch{\"a}fer},\ and\ \citenamefont
  {Ropers}}]{Feist2017}%
  \BibitemOpen
  \bibfield  {author} {\bibinfo {author} {\bibfnamefont {A.}~\bibnamefont
  {Feist}}, \bibinfo {author} {\bibfnamefont {N.}~\bibnamefont {Bach}},
  \bibinfo {author} {\bibfnamefont {N.}~\bibnamefont {{Rubiano da Silva}}},
  \bibinfo {author} {\bibfnamefont {T.}~\bibnamefont {Danz}}, \bibinfo {author}
  {\bibfnamefont {M.}~\bibnamefont {M{\"o}ller}}, \bibinfo {author}
  {\bibfnamefont {K.~E.}\ \bibnamefont {Priebe}}, \bibinfo {author}
  {\bibfnamefont {T.}~\bibnamefont {Domr{\"o}se}}, \bibinfo {author}
  {\bibfnamefont {J.~G.}\ \bibnamefont {Gatzmann}}, \bibinfo {author}
  {\bibfnamefont {S.}~\bibnamefont {Rost}}, \bibinfo {author} {\bibfnamefont
  {J.}~\bibnamefont {Schauss}}, \bibinfo {author} {\bibfnamefont
  {S.}~\bibnamefont {Strauch}}, \bibinfo {author} {\bibfnamefont
  {R.}~\bibnamefont {Bormann}}, \bibinfo {author} {\bibfnamefont
  {M.}~\bibnamefont {Sivis}}, \bibinfo {author} {\bibfnamefont
  {S.}~\bibnamefont {Sch{\"a}fer}},\ and\ \bibinfo {author} {\bibfnamefont
  {C.}~\bibnamefont {Ropers}},\ }\href {https://doi.org/10/gbmsq2} {\bibfield
  {journal} {\bibinfo  {journal} {Ultramicroscopy}\ }\textbf {\bibinfo {volume}
  {176}},\ \bibinfo {pages} {63} (\bibinfo {year} {2017})}\BibitemShut
  {NoStop}%
\bibitem [{\citenamefont {Bach}\ \emph {et~al.}(2019)\citenamefont {Bach},
  \citenamefont {Domr{\"o}se}, \citenamefont {Feist}, \citenamefont {Rittmann},
  \citenamefont {Strauch}, \citenamefont {Ropers},\ and\ \citenamefont
  {Sch{\"a}fer}}]{Bach2019}%
  \BibitemOpen
  \bibfield  {author} {\bibinfo {author} {\bibfnamefont {N.}~\bibnamefont
  {Bach}}, \bibinfo {author} {\bibfnamefont {T.}~\bibnamefont {Domr{\"o}se}},
  \bibinfo {author} {\bibfnamefont {A.}~\bibnamefont {Feist}}, \bibinfo
  {author} {\bibfnamefont {T.}~\bibnamefont {Rittmann}}, \bibinfo {author}
  {\bibfnamefont {S.}~\bibnamefont {Strauch}}, \bibinfo {author} {\bibfnamefont
  {C.}~\bibnamefont {Ropers}},\ and\ \bibinfo {author} {\bibfnamefont
  {S.}~\bibnamefont {Sch{\"a}fer}},\ }\href {https://doi.org/10.1063/1.5066093}
  {\bibfield  {journal} {\bibinfo  {journal} {Structural Dynamics}\ }\textbf
  {\bibinfo {volume} {6}},\ \bibinfo {pages} {014301} (\bibinfo {year}
  {2019})}\BibitemShut {NoStop}%
\bibitem [{\citenamefont {Hofmann}(2017)}]{Hofmann2017}%
  \BibitemOpen
  \bibfield  {author} {\bibinfo {author} {\bibfnamefont {I.}~\bibnamefont
  {Hofmann}},\ }\href {https://doi.org/10.1007/978-3-319-62157-9} {\emph
  {\bibinfo {title} {Space {{Charge Physics}} for {{Particle Accelerators}}}}}\
  (\bibinfo  {publisher} {{Springer Cham}},\ \bibinfo {year}
  {2017})\BibitemShut {NoStop}%
\bibitem [{\citenamefont {Emma}\ \emph {et~al.}(2006)\citenamefont {Emma},
  \citenamefont {Huang}, \citenamefont {Kim},\ and\ \citenamefont
  {Piot}}]{Emma2006}%
  \BibitemOpen
  \bibfield  {author} {\bibinfo {author} {\bibfnamefont {P.}~\bibnamefont
  {Emma}}, \bibinfo {author} {\bibfnamefont {Z.}~\bibnamefont {Huang}},
  \bibinfo {author} {\bibfnamefont {K.-J.}\ \bibnamefont {Kim}},\ and\ \bibinfo
  {author} {\bibfnamefont {P.}~\bibnamefont {Piot}},\ }\href
  {https://doi.org/10.1103/PhysRevSTAB.9.100702} {\bibfield  {journal}
  {\bibinfo  {journal} {Physical Review Special Topics - Accelerators and
  Beams}\ }\textbf {\bibinfo {volume} {9}},\ \bibinfo {pages} {100702}
  (\bibinfo {year} {2006})}\BibitemShut {NoStop}%
\bibitem [{\citenamefont {Kiesel}\ \emph {et~al.}(2002)\citenamefont {Kiesel},
  \citenamefont {Renz},\ and\ \citenamefont {Hasselbach}}]{Kiesel2002}%
  \BibitemOpen
  \bibfield  {author} {\bibinfo {author} {\bibfnamefont {H.}~\bibnamefont
  {Kiesel}}, \bibinfo {author} {\bibfnamefont {A.}~\bibnamefont {Renz}},\ and\
  \bibinfo {author} {\bibfnamefont {F.}~\bibnamefont {Hasselbach}},\ }\href
  {https://doi.org/10.1038/nature00911} {\bibfield  {journal} {\bibinfo
  {journal} {Nature}\ }\textbf {\bibinfo {volume} {418}},\ \bibinfo {pages}
  {392} (\bibinfo {year} {2002})}\BibitemShut {NoStop}%
\bibitem [{\citenamefont {Kuwahara}\ \emph {et~al.}(2021)\citenamefont
  {Kuwahara}, \citenamefont {Yoshida}, \citenamefont {Nagata}, \citenamefont
  {Nakakura}, \citenamefont {Furui}, \citenamefont {Ishida}, \citenamefont
  {Saitoh}, \citenamefont {Ujihara},\ and\ \citenamefont
  {Tanaka}}]{Kuwahara2021}%
  \BibitemOpen
  \bibfield  {author} {\bibinfo {author} {\bibfnamefont {M.}~\bibnamefont
  {Kuwahara}}, \bibinfo {author} {\bibfnamefont {Y.}~\bibnamefont {Yoshida}},
  \bibinfo {author} {\bibfnamefont {W.}~\bibnamefont {Nagata}}, \bibinfo
  {author} {\bibfnamefont {K.}~\bibnamefont {Nakakura}}, \bibinfo {author}
  {\bibfnamefont {M.}~\bibnamefont {Furui}}, \bibinfo {author} {\bibfnamefont
  {T.}~\bibnamefont {Ishida}}, \bibinfo {author} {\bibfnamefont
  {K.}~\bibnamefont {Saitoh}}, \bibinfo {author} {\bibfnamefont
  {T.}~\bibnamefont {Ujihara}},\ and\ \bibinfo {author} {\bibfnamefont
  {N.}~\bibnamefont {Tanaka}},\ }\href
  {https://doi.org/10.1103/PhysRevLett.126.125501} {\bibfield  {journal}
  {\bibinfo  {journal} {Physical Review Letters}\ }\textbf {\bibinfo {volume}
  {126}},\ \bibinfo {pages} {125501} (\bibinfo {year} {2021})}\BibitemShut
  {NoStop}%
\bibitem [{\citenamefont {Keramati}\ \emph {et~al.}(2021)\citenamefont
  {Keramati}, \citenamefont {Brunner}, \citenamefont {Gay},\ and\ \citenamefont
  {Batelaan}}]{Keramati2021}%
  \BibitemOpen
  \bibfield  {author} {\bibinfo {author} {\bibfnamefont {S.}~\bibnamefont
  {Keramati}}, \bibinfo {author} {\bibfnamefont {W.}~\bibnamefont {Brunner}},
  \bibinfo {author} {\bibfnamefont {T.~J.}\ \bibnamefont {Gay}},\ and\ \bibinfo
  {author} {\bibfnamefont {H.}~\bibnamefont {Batelaan}},\ }\href
  {https://doi.org/10.1103/physrevlett.127.180602} {\bibfield  {journal}
  {\bibinfo  {journal} {Physical Review Letters}\ }\textbf {\bibinfo {volume}
  {127}},\ \bibinfo {pages} {180602} (\bibinfo {year} {2021})}\BibitemShut
  {NoStop}%
\bibitem [{\citenamefont {Baym}\ and\ \citenamefont {Shen}(2014)}]{Baym2014}%
  \BibitemOpen
  \bibfield  {author} {\bibinfo {author} {\bibfnamefont {G.}~\bibnamefont
  {Baym}}\ and\ \bibinfo {author} {\bibfnamefont {K.}~\bibnamefont {Shen}},\
  }\bibinfo {title} {Hanbury {{Brown}}\textendash{{Twiss Interferometry}} with
  {{Electrons}}: {{Coulomb}} vs. {{Quantum Statistics}}},\ in\ \href
  {https://doi.org/10.1142/9789814472906_0024} {\emph {\bibinfo {booktitle} {In
  {{Memory}} of {{Akira Tonomura}}}}}\ (\bibinfo  {publisher} {{WORLD
  SCIENTIFIC}},\ \bibinfo {year} {2014})\ pp.\ \bibinfo {pages}
  {201--210}\BibitemShut {NoStop}%
\bibitem [{\citenamefont {Kodama}\ and\ \citenamefont
  {Osakabe}(2019)}]{Kodama2019}%
  \BibitemOpen
  \bibfield  {author} {\bibinfo {author} {\bibfnamefont {T.}~\bibnamefont
  {Kodama}}\ and\ \bibinfo {author} {\bibfnamefont {N.}~\bibnamefont
  {Osakabe}},\ }\href {https://doi.org/10.1093/jmicro/dfy129} {\bibfield
  {journal} {\bibinfo  {journal} {Microscopy}\ }\textbf {\bibinfo {volume}
  {68}},\ \bibinfo {pages} {133} (\bibinfo {year} {2019})}\BibitemShut
  {NoStop}%
\bibitem [{\citenamefont {Hommelhoff}\ \emph {et~al.}(2006)\citenamefont
  {Hommelhoff}, \citenamefont {Sortais}, \citenamefont {{Aghajani-Talesh}},\
  and\ \citenamefont {Kasevich}}]{Hommelhoff2006}%
  \BibitemOpen
  \bibfield  {author} {\bibinfo {author} {\bibfnamefont {P.}~\bibnamefont
  {Hommelhoff}}, \bibinfo {author} {\bibfnamefont {Y.}~\bibnamefont {Sortais}},
  \bibinfo {author} {\bibfnamefont {A.}~\bibnamefont {{Aghajani-Talesh}}},\
  and\ \bibinfo {author} {\bibfnamefont {M.~A.}\ \bibnamefont {Kasevich}},\
  }\href {https://doi.org/10.1103/PhysRevLett.96.077401} {\bibfield  {journal}
  {\bibinfo  {journal} {Physical Review Letters}\ }\textbf {\bibinfo {volume}
  {96}},\ \bibinfo {pages} {077401} (\bibinfo {year} {2006})}\BibitemShut
  {NoStop}%
\bibitem [{\citenamefont {Ropers}\ \emph {et~al.}(2007)\citenamefont {Ropers},
  \citenamefont {Solli}, \citenamefont {Schulz}, \citenamefont {Lienau},\ and\
  \citenamefont {Elsaesser}}]{Ropers2007}%
  \BibitemOpen
  \bibfield  {author} {\bibinfo {author} {\bibfnamefont {C.}~\bibnamefont
  {Ropers}}, \bibinfo {author} {\bibfnamefont {D.~R.}\ \bibnamefont {Solli}},
  \bibinfo {author} {\bibfnamefont {C.~P.}\ \bibnamefont {Schulz}}, \bibinfo
  {author} {\bibfnamefont {C.}~\bibnamefont {Lienau}},\ and\ \bibinfo {author}
  {\bibfnamefont {T.}~\bibnamefont {Elsaesser}},\ }\href
  {https://doi.org/10/brxccv} {\bibfield  {journal} {\bibinfo  {journal}
  {Physical Review Letters}\ }\textbf {\bibinfo {volume} {98}},\ \bibinfo
  {pages} {043907} (\bibinfo {year} {2007})}\BibitemShut {NoStop}%
\bibitem [{\citenamefont {Barwick}\ \emph {et~al.}(2007)\citenamefont
  {Barwick}, \citenamefont {Corder}, \citenamefont {Strohaber}, \citenamefont
  {{Chandler-Smith}}, \citenamefont {Uiterwaal},\ and\ \citenamefont
  {Batelaan}}]{Barwick2007}%
  \BibitemOpen
  \bibfield  {author} {\bibinfo {author} {\bibfnamefont {B.}~\bibnamefont
  {Barwick}}, \bibinfo {author} {\bibfnamefont {C.}~\bibnamefont {Corder}},
  \bibinfo {author} {\bibfnamefont {J.}~\bibnamefont {Strohaber}}, \bibinfo
  {author} {\bibfnamefont {N.}~\bibnamefont {{Chandler-Smith}}}, \bibinfo
  {author} {\bibfnamefont {C.}~\bibnamefont {Uiterwaal}},\ and\ \bibinfo
  {author} {\bibfnamefont {H.}~\bibnamefont {Batelaan}},\ }\href
  {https://doi.org/10.1088/1367-2630/9/5/142} {\bibfield  {journal} {\bibinfo
  {journal} {New Journal of Physics}\ }\textbf {\bibinfo {volume} {9}},\
  \bibinfo {pages} {142} (\bibinfo {year} {2007})}\BibitemShut {NoStop}%
\bibitem [{\citenamefont {Kr{\"u}ger}\ \emph {et~al.}(2011)\citenamefont
  {Kr{\"u}ger}, \citenamefont {Schenk},\ and\ \citenamefont
  {Hommelhoff}}]{Kruger2011}%
  \BibitemOpen
  \bibfield  {author} {\bibinfo {author} {\bibfnamefont {M.}~\bibnamefont
  {Kr{\"u}ger}}, \bibinfo {author} {\bibfnamefont {M.}~\bibnamefont {Schenk}},\
  and\ \bibinfo {author} {\bibfnamefont {P.}~\bibnamefont {Hommelhoff}},\
  }\href {https://doi.org/10/ff9khf} {\bibfield  {journal} {\bibinfo  {journal}
  {Nature}\ }\textbf {\bibinfo {volume} {475}},\ \bibinfo {pages} {78}
  (\bibinfo {year} {2011})}\BibitemShut {NoStop}%
\bibitem [{\citenamefont {Herink}\ \emph {et~al.}(2012)\citenamefont {Herink},
  \citenamefont {Solli}, \citenamefont {Gulde},\ and\ \citenamefont
  {Ropers}}]{Herink2012}%
  \BibitemOpen
  \bibfield  {author} {\bibinfo {author} {\bibfnamefont {G.}~\bibnamefont
  {Herink}}, \bibinfo {author} {\bibfnamefont {D.~R.}\ \bibnamefont {Solli}},
  \bibinfo {author} {\bibfnamefont {M.}~\bibnamefont {Gulde}},\ and\ \bibinfo
  {author} {\bibfnamefont {C.}~\bibnamefont {Ropers}},\ }\href
  {https://doi.org/10/ggnzn5} {\bibfield  {journal} {\bibinfo  {journal}
  {Nature}\ }\textbf {\bibinfo {volume} {483}},\ \bibinfo {pages} {190}
  (\bibinfo {year} {2012})}\BibitemShut {NoStop}%
\bibitem [{\citenamefont {Ciappina}\ \emph {et~al.}(2017)\citenamefont
  {Ciappina}, \citenamefont {{P{\'e}rez-Hern{\'a}ndez}}, \citenamefont
  {Landsman}, \citenamefont {Okell}, \citenamefont {Zherebtsov}, \citenamefont
  {F{\"o}rg}, \citenamefont {Sch{\"o}tz}, \citenamefont {Seiffert},
  \citenamefont {Fennel}, \citenamefont {Shaaran}, \citenamefont {Zimmermann},
  \citenamefont {Chac{\'o}n}, \citenamefont {Guichard}, \citenamefont
  {Za{\"i}r}, \citenamefont {Tisch}, \citenamefont {Marangos}, \citenamefont
  {Witting}, \citenamefont {Braun}, \citenamefont {Maier}, \citenamefont
  {Roso}, \citenamefont {Kr{\"u}ger}, \citenamefont {Hommelhoff}, \citenamefont
  {Kling}, \citenamefont {Krausz},\ and\ \citenamefont
  {Lewenstein}}]{Ciappina2017}%
  \BibitemOpen
  \bibfield  {author} {\bibinfo {author} {\bibfnamefont {M.~F.}\ \bibnamefont
  {Ciappina}}, \bibinfo {author} {\bibfnamefont {J.~A.}\ \bibnamefont
  {{P{\'e}rez-Hern{\'a}ndez}}}, \bibinfo {author} {\bibfnamefont {A.~S.}\
  \bibnamefont {Landsman}}, \bibinfo {author} {\bibfnamefont {W.~A.}\
  \bibnamefont {Okell}}, \bibinfo {author} {\bibfnamefont {S.}~\bibnamefont
  {Zherebtsov}}, \bibinfo {author} {\bibfnamefont {B.}~\bibnamefont
  {F{\"o}rg}}, \bibinfo {author} {\bibfnamefont {J.}~\bibnamefont
  {Sch{\"o}tz}}, \bibinfo {author} {\bibfnamefont {L.}~\bibnamefont
  {Seiffert}}, \bibinfo {author} {\bibfnamefont {T.}~\bibnamefont {Fennel}},
  \bibinfo {author} {\bibfnamefont {T.}~\bibnamefont {Shaaran}}, \bibinfo
  {author} {\bibfnamefont {T.}~\bibnamefont {Zimmermann}}, \bibinfo {author}
  {\bibfnamefont {A.}~\bibnamefont {Chac{\'o}n}}, \bibinfo {author}
  {\bibfnamefont {R.}~\bibnamefont {Guichard}}, \bibinfo {author}
  {\bibfnamefont {A.}~\bibnamefont {Za{\"i}r}}, \bibinfo {author}
  {\bibfnamefont {J.~W.~G.}\ \bibnamefont {Tisch}}, \bibinfo {author}
  {\bibfnamefont {J.~P.}\ \bibnamefont {Marangos}}, \bibinfo {author}
  {\bibfnamefont {T.}~\bibnamefont {Witting}}, \bibinfo {author} {\bibfnamefont
  {A.}~\bibnamefont {Braun}}, \bibinfo {author} {\bibfnamefont {S.~A.}\
  \bibnamefont {Maier}}, \bibinfo {author} {\bibfnamefont {L.}~\bibnamefont
  {Roso}}, \bibinfo {author} {\bibfnamefont {M.}~\bibnamefont {Kr{\"u}ger}},
  \bibinfo {author} {\bibfnamefont {P.}~\bibnamefont {Hommelhoff}}, \bibinfo
  {author} {\bibfnamefont {M.~F.}\ \bibnamefont {Kling}}, \bibinfo {author}
  {\bibfnamefont {F.}~\bibnamefont {Krausz}},\ and\ \bibinfo {author}
  {\bibfnamefont {M.}~\bibnamefont {Lewenstein}},\ }\href
  {https://doi.org/10.1088/1361-6633/aa574e} {\bibfield  {journal} {\bibinfo
  {journal} {Reports on Progress in Physics}\ }\textbf {\bibinfo {volume}
  {80}},\ \bibinfo {pages} {054401} (\bibinfo {year} {2017})}\BibitemShut
  {NoStop}%
\bibitem [{\citenamefont {Seiffert}\ \emph {et~al.}(2018)\citenamefont
  {Seiffert}, \citenamefont {Paschen}, \citenamefont {Hommelhoff},\ and\
  \citenamefont {Fennel}}]{Seiffert2018}%
  \BibitemOpen
  \bibfield  {author} {\bibinfo {author} {\bibfnamefont {L.}~\bibnamefont
  {Seiffert}}, \bibinfo {author} {\bibfnamefont {T.}~\bibnamefont {Paschen}},
  \bibinfo {author} {\bibfnamefont {P.}~\bibnamefont {Hommelhoff}},\ and\
  \bibinfo {author} {\bibfnamefont {T.}~\bibnamefont {Fennel}},\ }\href
  {https://doi.org/10.1088/1361-6455/aac34f} {\bibfield  {journal} {\bibinfo
  {journal} {Journal of Physics B: Atomic, Molecular and Optical Physics}\
  }\textbf {\bibinfo {volume} {51}},\ \bibinfo {pages} {134001} (\bibinfo
  {year} {2018})}\BibitemShut {NoStop}%
\bibitem [{\citenamefont {Dombi}\ \emph {et~al.}(2020)\citenamefont {Dombi},
  \citenamefont {P{\'a}pa}, \citenamefont {Vogelsang}, \citenamefont {Yalunin},
  \citenamefont {Sivis}, \citenamefont {Herink}, \citenamefont {Sch{\"a}fer},
  \citenamefont {Gro{\ss}}, \citenamefont {Ropers},\ and\ \citenamefont
  {Lienau}}]{Dombi2020}%
  \BibitemOpen
  \bibfield  {author} {\bibinfo {author} {\bibfnamefont {P.}~\bibnamefont
  {Dombi}}, \bibinfo {author} {\bibfnamefont {Z.}~\bibnamefont {P{\'a}pa}},
  \bibinfo {author} {\bibfnamefont {J.}~\bibnamefont {Vogelsang}}, \bibinfo
  {author} {\bibfnamefont {S.~V.}\ \bibnamefont {Yalunin}}, \bibinfo {author}
  {\bibfnamefont {M.}~\bibnamefont {Sivis}}, \bibinfo {author} {\bibfnamefont
  {G.}~\bibnamefont {Herink}}, \bibinfo {author} {\bibfnamefont
  {S.}~\bibnamefont {Sch{\"a}fer}}, \bibinfo {author} {\bibfnamefont
  {P.}~\bibnamefont {Gro{\ss}}}, \bibinfo {author} {\bibfnamefont
  {C.}~\bibnamefont {Ropers}},\ and\ \bibinfo {author} {\bibfnamefont
  {C.}~\bibnamefont {Lienau}},\ }\href
  {https://doi.org/10.1103/RevModPhys.92.025003} {\bibfield  {journal}
  {\bibinfo  {journal} {Reviews of Modern Physics}\ }\textbf {\bibinfo {volume}
  {92}},\ \bibinfo {pages} {025003} (\bibinfo {year} {2020})}\BibitemShut
  {NoStop}%
\bibitem [{\citenamefont {Houdellier}\ \emph {et~al.}(2018)\citenamefont
  {Houdellier}, \citenamefont {Caruso}, \citenamefont {Weber}, \citenamefont
  {Kociak},\ and\ \citenamefont {Arbouet}}]{Houdellier2018}%
  \BibitemOpen
  \bibfield  {author} {\bibinfo {author} {\bibfnamefont {F.}~\bibnamefont
  {Houdellier}}, \bibinfo {author} {\bibfnamefont {G.}~\bibnamefont {Caruso}},
  \bibinfo {author} {\bibfnamefont {S.}~\bibnamefont {Weber}}, \bibinfo
  {author} {\bibfnamefont {M.}~\bibnamefont {Kociak}},\ and\ \bibinfo {author}
  {\bibfnamefont {A.}~\bibnamefont {Arbouet}},\ }\href
  {https://doi.org/10/gc6pjc} {\bibfield  {journal} {\bibinfo  {journal}
  {Ultramicroscopy}\ }\textbf {\bibinfo {volume} {186}},\ \bibinfo {pages}
  {128} (\bibinfo {year} {2018})}\BibitemShut {NoStop}%
\bibitem [{\citenamefont {Yanagisawa}\ \emph {et~al.}(2016)\citenamefont
  {Yanagisawa}, \citenamefont {Schnepp}, \citenamefont {Hafner}, \citenamefont
  {Hengsberger}, \citenamefont {Kim}, \citenamefont {Kling}, \citenamefont
  {Landsman}, \citenamefont {Gallmann},\ and\ \citenamefont
  {Osterwalder}}]{Yanagisawa2016}%
  \BibitemOpen
  \bibfield  {author} {\bibinfo {author} {\bibfnamefont {H.}~\bibnamefont
  {Yanagisawa}}, \bibinfo {author} {\bibfnamefont {S.}~\bibnamefont {Schnepp}},
  \bibinfo {author} {\bibfnamefont {C.}~\bibnamefont {Hafner}}, \bibinfo
  {author} {\bibfnamefont {M.}~\bibnamefont {Hengsberger}}, \bibinfo {author}
  {\bibfnamefont {D.~E.}\ \bibnamefont {Kim}}, \bibinfo {author} {\bibfnamefont
  {M.~F.}\ \bibnamefont {Kling}}, \bibinfo {author} {\bibfnamefont
  {A.}~\bibnamefont {Landsman}}, \bibinfo {author} {\bibfnamefont
  {L.}~\bibnamefont {Gallmann}},\ and\ \bibinfo {author} {\bibfnamefont
  {J.}~\bibnamefont {Osterwalder}},\ }\href {https://doi.org/10.1038/srep35877}
  {\bibfield  {journal} {\bibinfo  {journal} {Scientific Reports}\ }\textbf
  {\bibinfo {volume} {6}},\ \bibinfo {pages} {35877} (\bibinfo {year}
  {2016})}\BibitemShut {NoStop}%
\bibitem [{\citenamefont {Sch{\"o}tz}\ \emph {et~al.}(2021)\citenamefont
  {Sch{\"o}tz}, \citenamefont {Seiffert}, \citenamefont {Maliakkal},
  \citenamefont {Bl{\"o}chl}, \citenamefont {Zimin}, \citenamefont
  {Rosenberger}, \citenamefont {Bergues}, \citenamefont {Hommelhoff},
  \citenamefont {Krausz}, \citenamefont {Fennel},\ and\ \citenamefont
  {Kling}}]{Schotz2021}%
  \BibitemOpen
  \bibfield  {author} {\bibinfo {author} {\bibfnamefont {J.}~\bibnamefont
  {Sch{\"o}tz}}, \bibinfo {author} {\bibfnamefont {L.}~\bibnamefont
  {Seiffert}}, \bibinfo {author} {\bibfnamefont {A.}~\bibnamefont {Maliakkal}},
  \bibinfo {author} {\bibfnamefont {J.}~\bibnamefont {Bl{\"o}chl}}, \bibinfo
  {author} {\bibfnamefont {D.}~\bibnamefont {Zimin}}, \bibinfo {author}
  {\bibfnamefont {P.}~\bibnamefont {Rosenberger}}, \bibinfo {author}
  {\bibfnamefont {B.}~\bibnamefont {Bergues}}, \bibinfo {author} {\bibfnamefont
  {P.}~\bibnamefont {Hommelhoff}}, \bibinfo {author} {\bibfnamefont
  {F.}~\bibnamefont {Krausz}}, \bibinfo {author} {\bibfnamefont
  {T.}~\bibnamefont {Fennel}},\ and\ \bibinfo {author} {\bibfnamefont {M.~F.}\
  \bibnamefont {Kling}},\ }\href {https://doi.org/10.1515/nanoph-2021-0276}
  {\bibfield  {journal} {\bibinfo  {journal} {Nanophotonics}\ }\textbf
  {\bibinfo {volume} {10}},\ \bibinfo {pages} {3769} (\bibinfo {year}
  {2021})}\BibitemShut {NoStop}%
\bibitem [{\citenamefont {Meier}\ and\ \citenamefont
  {Hommelhoff}(2022)}]{Meier2022a}%
  \BibitemOpen
  \bibfield  {author} {\bibinfo {author} {\bibfnamefont {S.}~\bibnamefont
  {Meier}}\ and\ \bibinfo {author} {\bibfnamefont {P.}~\bibnamefont
  {Hommelhoff}},\ }\href {https://doi.org/10.1021/acsphotonics.2c00839}
  {\bibfield  {journal} {\bibinfo  {journal} {ACS Photonics}\ }\textbf
  {\bibinfo {volume} {9}},\ \bibinfo {pages} {3083} (\bibinfo {year}
  {2022})}\BibitemShut {NoStop}%
\bibitem [{\citenamefont {Mandel}\ and\ \citenamefont
  {Wolf}(1995)}]{Mandel1995}%
  \BibitemOpen
  \bibfield  {author} {\bibinfo {author} {\bibfnamefont {L.}~\bibnamefont
  {Mandel}}\ and\ \bibinfo {author} {\bibfnamefont {E.}~\bibnamefont {Wolf}},\
  }\href {https://doi.org/10.1017/CBO9781139644105} {\emph {\bibinfo {title}
  {Optical {{Coherence}} and {{Quantum Optics}}}}},\ \bibinfo {edition} {1st}\
  ed.\ (\bibinfo  {publisher} {{Cambridge University Press}},\ \bibinfo {year}
  {1995})\BibitemShut {NoStop}%
\bibitem [{\citenamefont {Kodama}\ \emph {et~al.}(2011)\citenamefont {Kodama},
  \citenamefont {Osakabe},\ and\ \citenamefont {Tonomura}}]{Kodama2011}%
  \BibitemOpen
  \bibfield  {author} {\bibinfo {author} {\bibfnamefont {T.}~\bibnamefont
  {Kodama}}, \bibinfo {author} {\bibfnamefont {N.}~\bibnamefont {Osakabe}},\
  and\ \bibinfo {author} {\bibfnamefont {A.}~\bibnamefont {Tonomura}},\ }\href
  {https://doi.org/10.1103/PhysRevA.83.063616} {\bibfield  {journal} {\bibinfo
  {journal} {Physical Review A}\ }\textbf {\bibinfo {volume} {83}},\ \bibinfo
  {pages} {063616} (\bibinfo {year} {2011})}\BibitemShut {NoStop}%
\bibitem [{\citenamefont {D{\"o}rner}\ \emph {et~al.}(2000)\citenamefont
  {D{\"o}rner}, \citenamefont {Mergel}, \citenamefont {Jagutzki}, \citenamefont
  {Spielberger}, \citenamefont {Ullrich}, \citenamefont {Moshammer},\ and\
  \citenamefont {{Schmidt-B{\"o}cking}}}]{Dorner2000}%
  \BibitemOpen
  \bibfield  {author} {\bibinfo {author} {\bibfnamefont {R.}~\bibnamefont
  {D{\"o}rner}}, \bibinfo {author} {\bibfnamefont {V.}~\bibnamefont {Mergel}},
  \bibinfo {author} {\bibfnamefont {O.}~\bibnamefont {Jagutzki}}, \bibinfo
  {author} {\bibfnamefont {L.}~\bibnamefont {Spielberger}}, \bibinfo {author}
  {\bibfnamefont {J.}~\bibnamefont {Ullrich}}, \bibinfo {author} {\bibfnamefont
  {R.}~\bibnamefont {Moshammer}},\ and\ \bibinfo {author} {\bibfnamefont
  {H.}~\bibnamefont {{Schmidt-B{\"o}cking}}},\ }\href
  {https://doi.org/10.1016/S0370-1573(99)00109-X} {\bibfield  {journal}
  {\bibinfo  {journal} {Physics Reports}\ }\textbf {\bibinfo {volume} {330}},\
  \bibinfo {pages} {95} (\bibinfo {year} {2000})}\BibitemShut {NoStop}%
\bibitem [{\citenamefont {Ullrich}\ \emph {et~al.}(2003)\citenamefont
  {Ullrich}, \citenamefont {Moshammer}, \citenamefont {Dorn}, \citenamefont {{D
  rner}}, \citenamefont {Schmidt},\ and\ \citenamefont {{Schmidt-B
  cking}}}]{Ullrich2003}%
  \BibitemOpen
  \bibfield  {author} {\bibinfo {author} {\bibfnamefont {J.}~\bibnamefont
  {Ullrich}}, \bibinfo {author} {\bibfnamefont {R.}~\bibnamefont {Moshammer}},
  \bibinfo {author} {\bibfnamefont {A.}~\bibnamefont {Dorn}}, \bibinfo {author}
  {\bibfnamefont {R.}~\bibnamefont {{D rner}}}, \bibinfo {author}
  {\bibfnamefont {L.~P.~H.}\ \bibnamefont {Schmidt}},\ and\ \bibinfo {author}
  {\bibfnamefont {H.}~\bibnamefont {{Schmidt-B cking}}},\ }\href
  {https://doi.org/10.1088/0034-4885/66/9/203} {\bibfield  {journal} {\bibinfo
  {journal} {Reports on Progress in Physics}\ }\textbf {\bibinfo {volume}
  {66}},\ \bibinfo {pages} {1463} (\bibinfo {year} {2003})}\BibitemShut
  {NoStop}%
\bibitem [{\citenamefont {{Mu{\~n}oz-Navia}}\ \emph {et~al.}(2009)\citenamefont
  {{Mu{\~n}oz-Navia}}, \citenamefont {Winkler}, \citenamefont {Patel},
  \citenamefont {Birke}, \citenamefont {Schumann},\ and\ \citenamefont
  {Kirschner}}]{Munoz-Navia2009}%
  \BibitemOpen
  \bibfield  {author} {\bibinfo {author} {\bibfnamefont {M.}~\bibnamefont
  {{Mu{\~n}oz-Navia}}}, \bibinfo {author} {\bibfnamefont {C.}~\bibnamefont
  {Winkler}}, \bibinfo {author} {\bibfnamefont {R.}~\bibnamefont {Patel}},
  \bibinfo {author} {\bibfnamefont {M.}~\bibnamefont {Birke}}, \bibinfo
  {author} {\bibfnamefont {F.~O.}\ \bibnamefont {Schumann}},\ and\ \bibinfo
  {author} {\bibfnamefont {J.}~\bibnamefont {Kirschner}},\ }\href
  {https://doi.org/10.1088/0953-8984/21/35/355003} {\bibfield  {journal}
  {\bibinfo  {journal} {Journal of Physics: Condensed Matter}\ }\textbf
  {\bibinfo {volume} {21}},\ \bibinfo {pages} {355003} (\bibinfo {year}
  {2009})}\BibitemShut {NoStop}%
\bibitem [{\citenamefont {{van Riessen}}\ \emph {et~al.}(2010)\citenamefont
  {{van Riessen}}, \citenamefont {Wei}, \citenamefont {Dhaka}, \citenamefont
  {Winkler}, \citenamefont {Schumann},\ and\ \citenamefont
  {Kirschner}}]{vanRiessen2010}%
  \BibitemOpen
  \bibfield  {author} {\bibinfo {author} {\bibfnamefont {G.}~\bibnamefont {{van
  Riessen}}}, \bibinfo {author} {\bibfnamefont {Z.}~\bibnamefont {Wei}},
  \bibinfo {author} {\bibfnamefont {R.~S.}\ \bibnamefont {Dhaka}}, \bibinfo
  {author} {\bibfnamefont {C.}~\bibnamefont {Winkler}}, \bibinfo {author}
  {\bibfnamefont {F.~O.}\ \bibnamefont {Schumann}},\ and\ \bibinfo {author}
  {\bibfnamefont {J.}~\bibnamefont {Kirschner}},\ }\href
  {https://doi.org/10.1088/0953-8984/22/9/092201} {\bibfield  {journal}
  {\bibinfo  {journal} {Journal of Physics: Condensed Matter}\ }\textbf
  {\bibinfo {volume} {22}},\ \bibinfo {pages} {092201} (\bibinfo {year}
  {2010})}\BibitemShut {NoStop}%
\bibitem [{\citenamefont {Tr{\"u}tzschler}\ \emph {et~al.}(2017)\citenamefont
  {Tr{\"u}tzschler}, \citenamefont {Huth}, \citenamefont {Chiang},
  \citenamefont {Kamrla}, \citenamefont {Schumann}, \citenamefont {Kirschner},\
  and\ \citenamefont {Widdra}}]{Trutzschler2017}%
  \BibitemOpen
  \bibfield  {author} {\bibinfo {author} {\bibfnamefont {A.}~\bibnamefont
  {Tr{\"u}tzschler}}, \bibinfo {author} {\bibfnamefont {M.}~\bibnamefont
  {Huth}}, \bibinfo {author} {\bibfnamefont {C.-T.}\ \bibnamefont {Chiang}},
  \bibinfo {author} {\bibfnamefont {R.}~\bibnamefont {Kamrla}}, \bibinfo
  {author} {\bibfnamefont {F.~O.}\ \bibnamefont {Schumann}}, \bibinfo {author}
  {\bibfnamefont {J.}~\bibnamefont {Kirschner}},\ and\ \bibinfo {author}
  {\bibfnamefont {W.}~\bibnamefont {Widdra}},\ }\href
  {https://doi.org/10.1103/PhysRevLett.118.136401} {\bibfield  {journal}
  {\bibinfo  {journal} {Physical Review Letters}\ }\textbf {\bibinfo {volume}
  {118}},\ \bibinfo {pages} {136401} (\bibinfo {year} {2017})}\BibitemShut
  {NoStop}%
\bibitem [{\citenamefont {Larochelle}\ \emph {et~al.}(1998)\citenamefont
  {Larochelle}, \citenamefont {Talebpour},\ and\ \citenamefont
  {Chin}}]{Larochelle1998}%
  \BibitemOpen
  \bibfield  {author} {\bibinfo {author} {\bibfnamefont {S.}~\bibnamefont
  {Larochelle}}, \bibinfo {author} {\bibfnamefont {A.}~\bibnamefont
  {Talebpour}},\ and\ \bibinfo {author} {\bibfnamefont {S.~L.}\ \bibnamefont
  {Chin}},\ }\href {https://doi.org/10.1088/0953-4075/31/6/008} {\bibfield
  {journal} {\bibinfo  {journal} {Journal of Physics B: Atomic, Molecular and
  Optical Physics}\ }\textbf {\bibinfo {volume} {31}},\ \bibinfo {pages} {1201}
  (\bibinfo {year} {1998})}\BibitemShut {NoStop}%
\bibitem [{\citenamefont {Becker}\ \emph {et~al.}(2012)\citenamefont {Becker},
  \citenamefont {Liu}, \citenamefont {Ho},\ and\ \citenamefont
  {Eberly}}]{Becker2012}%
  \BibitemOpen
  \bibfield  {author} {\bibinfo {author} {\bibfnamefont {W.}~\bibnamefont
  {Becker}}, \bibinfo {author} {\bibfnamefont {X.}~\bibnamefont {Liu}},
  \bibinfo {author} {\bibfnamefont {P.~J.}\ \bibnamefont {Ho}},\ and\ \bibinfo
  {author} {\bibfnamefont {J.~H.}\ \bibnamefont {Eberly}},\ }\href
  {https://doi.org/10.1103/RevModPhys.84.1011} {\bibfield  {journal} {\bibinfo
  {journal} {Reviews of Modern Physics}\ }\textbf {\bibinfo {volume} {84}},\
  \bibinfo {pages} {1011} (\bibinfo {year} {2012})}\BibitemShut {NoStop}%
\bibitem [{\citenamefont {Jannis}\ \emph {et~al.}(2019)\citenamefont {Jannis},
  \citenamefont {{M{\"u}ller-Caspary}}, \citenamefont {B{\'e}ch{\'e}},
  \citenamefont {Oelsner},\ and\ \citenamefont {Verbeeck}}]{Jannis2019}%
  \BibitemOpen
  \bibfield  {author} {\bibinfo {author} {\bibfnamefont {D.}~\bibnamefont
  {Jannis}}, \bibinfo {author} {\bibfnamefont {K.}~\bibnamefont
  {{M{\"u}ller-Caspary}}}, \bibinfo {author} {\bibfnamefont {A.}~\bibnamefont
  {B{\'e}ch{\'e}}}, \bibinfo {author} {\bibfnamefont {A.}~\bibnamefont
  {Oelsner}},\ and\ \bibinfo {author} {\bibfnamefont {J.}~\bibnamefont
  {Verbeeck}},\ }\href {https://doi.org/10.1063/1.5092945} {\bibfield
  {journal} {\bibinfo  {journal} {Applied Physics Letters}\ }\textbf {\bibinfo
  {volume} {114}},\ \bibinfo {pages} {143101} (\bibinfo {year}
  {2019})}\BibitemShut {NoStop}%
\bibitem [{\citenamefont {Varkentina}\ \emph {et~al.}(2022)\citenamefont
  {Varkentina}, \citenamefont {Auad}, \citenamefont {Woo}, \citenamefont
  {Zobelli}, \citenamefont {Bocher}, \citenamefont {Blazit}, \citenamefont
  {Li}, \citenamefont {Tenc{\'e}}, \citenamefont {Watanabe}, \citenamefont
  {Taniguchi}, \citenamefont {St{\'e}phan}, \citenamefont {Kociak},\ and\
  \citenamefont {Tizei}}]{Varkentina2022}%
  \BibitemOpen
  \bibfield  {author} {\bibinfo {author} {\bibfnamefont {N.}~\bibnamefont
  {Varkentina}}, \bibinfo {author} {\bibfnamefont {Y.}~\bibnamefont {Auad}},
  \bibinfo {author} {\bibfnamefont {S.~Y.}\ \bibnamefont {Woo}}, \bibinfo
  {author} {\bibfnamefont {A.}~\bibnamefont {Zobelli}}, \bibinfo {author}
  {\bibfnamefont {L.}~\bibnamefont {Bocher}}, \bibinfo {author} {\bibfnamefont
  {J.-D.}\ \bibnamefont {Blazit}}, \bibinfo {author} {\bibfnamefont
  {X.}~\bibnamefont {Li}}, \bibinfo {author} {\bibfnamefont {M.}~\bibnamefont
  {Tenc{\'e}}}, \bibinfo {author} {\bibfnamefont {K.}~\bibnamefont {Watanabe}},
  \bibinfo {author} {\bibfnamefont {T.}~\bibnamefont {Taniguchi}}, \bibinfo
  {author} {\bibfnamefont {O.}~\bibnamefont {St{\'e}phan}}, \bibinfo {author}
  {\bibfnamefont {M.}~\bibnamefont {Kociak}},\ and\ \bibinfo {author}
  {\bibfnamefont {L.~H.~G.}\ \bibnamefont {Tizei}},\ }\href
  {https://doi.org/10.1126/sciadv.abq4947} {\bibfield  {journal} {\bibinfo
  {journal} {Science Advances}\ }\textbf {\bibinfo {volume} {8}},\ \bibinfo
  {pages} {eabq4947} (\bibinfo {year} {2022})}\BibitemShut {NoStop}%
\bibitem [{\citenamefont {Feist}\ \emph {et~al.}(2022)\citenamefont {Feist},
  \citenamefont {Huang}, \citenamefont {Arend}, \citenamefont {Yang},
  \citenamefont {Henke}, \citenamefont {Raja}, \citenamefont {Kappert},
  \citenamefont {Wang}, \citenamefont {{Louren{\c c}o-Martins}}, \citenamefont
  {Qiu}, \citenamefont {Liu}, \citenamefont {Kfir}, \citenamefont
  {Kippenberg},\ and\ \citenamefont {Ropers}}]{Feist2022}%
  \BibitemOpen
  \bibfield  {author} {\bibinfo {author} {\bibfnamefont {A.}~\bibnamefont
  {Feist}}, \bibinfo {author} {\bibfnamefont {G.}~\bibnamefont {Huang}},
  \bibinfo {author} {\bibfnamefont {G.}~\bibnamefont {Arend}}, \bibinfo
  {author} {\bibfnamefont {Y.}~\bibnamefont {Yang}}, \bibinfo {author}
  {\bibfnamefont {J.-W.}\ \bibnamefont {Henke}}, \bibinfo {author}
  {\bibfnamefont {A.~S.}\ \bibnamefont {Raja}}, \bibinfo {author}
  {\bibfnamefont {F.~J.}\ \bibnamefont {Kappert}}, \bibinfo {author}
  {\bibfnamefont {R.~N.}\ \bibnamefont {Wang}}, \bibinfo {author}
  {\bibfnamefont {H.}~\bibnamefont {{Louren{\c c}o-Martins}}}, \bibinfo
  {author} {\bibfnamefont {Z.}~\bibnamefont {Qiu}}, \bibinfo {author}
  {\bibfnamefont {J.}~\bibnamefont {Liu}}, \bibinfo {author} {\bibfnamefont
  {O.}~\bibnamefont {Kfir}}, \bibinfo {author} {\bibfnamefont {T.~J.}\
  \bibnamefont {Kippenberg}},\ and\ \bibinfo {author} {\bibfnamefont
  {C.}~\bibnamefont {Ropers}},\ }\href
  {https://doi.org/10.1126/science.abo5037} {\bibfield  {journal} {\bibinfo
  {journal} {Science}\ }\textbf {\bibinfo {volume} {377}},\ \bibinfo {pages}
  {777} (\bibinfo {year} {2022})}\BibitemShut {NoStop}%
\bibitem [{\citenamefont {Kfir}\ \emph {et~al.}(2021)\citenamefont {Kfir},
  \citenamefont {Di~Giulio}, \citenamefont {{de Abajo}},\ and\ \citenamefont
  {Ropers}}]{Kfir2021}%
  \BibitemOpen
  \bibfield  {author} {\bibinfo {author} {\bibfnamefont {O.}~\bibnamefont
  {Kfir}}, \bibinfo {author} {\bibfnamefont {V.}~\bibnamefont {Di~Giulio}},
  \bibinfo {author} {\bibfnamefont {F.~J.~G.}\ \bibnamefont {{de Abajo}}},\
  and\ \bibinfo {author} {\bibfnamefont {C.}~\bibnamefont {Ropers}},\ }\href
  {https://doi.org/10.1126/sciadv.abf6380} {\bibfield  {journal} {\bibinfo
  {journal} {Science Advances}\ }\textbf {\bibinfo {volume} {7}},\ \bibinfo
  {pages} {eabf6380} (\bibinfo {year} {2021})}\BibitemShut {NoStop}%
\bibitem [{\citenamefont {Asban}\ and\ \citenamefont {{Garc{\'i}a de
  Abajo}}(2021)}]{Asban2021}%
  \BibitemOpen
  \bibfield  {author} {\bibinfo {author} {\bibfnamefont {S.}~\bibnamefont
  {Asban}}\ and\ \bibinfo {author} {\bibfnamefont {F.~J.}\ \bibnamefont
  {{Garc{\'i}a de Abajo}}},\ }\href
  {https://doi.org/10.1038/s41534-021-00376-4} {\bibfield  {journal} {\bibinfo
  {journal} {npj Quantum Information}\ }\textbf {\bibinfo {volume} {7}},\
  \bibinfo {pages} {42} (\bibinfo {year} {2021})}\BibitemShut {NoStop}%
\bibitem [{\citenamefont {Zhao}\ \emph {et~al.}(2021)\citenamefont {Zhao},
  \citenamefont {Sun},\ and\ \citenamefont {Fan}}]{Zhao2021}%
  \BibitemOpen
  \bibfield  {author} {\bibinfo {author} {\bibfnamefont {Z.}~\bibnamefont
  {Zhao}}, \bibinfo {author} {\bibfnamefont {X.-Q.}\ \bibnamefont {Sun}},\ and\
  \bibinfo {author} {\bibfnamefont {S.}~\bibnamefont {Fan}},\ }\href
  {https://doi.org/10.1103/physrevlett.126.233402} {\bibfield  {journal}
  {\bibinfo  {journal} {Physical Review Letters}\ }\textbf {\bibinfo {volume}
  {126}},\ \bibinfo {pages} {233402} (\bibinfo {year} {2021})}\BibitemShut
  {NoStop}%
\bibitem [{\citenamefont {Pan}\ and\ \citenamefont {Gover}(2019)}]{Pan2019}%
  \BibitemOpen
  \bibfield  {author} {\bibinfo {author} {\bibfnamefont {Y.}~\bibnamefont
  {Pan}}\ and\ \bibinfo {author} {\bibfnamefont {A.}~\bibnamefont {Gover}},\
  }\href {https://doi.org/10.1103/PhysRevA.99.052107} {\bibfield  {journal}
  {\bibinfo  {journal} {Physical Review A}\ }\textbf {\bibinfo {volume} {99}},\
  \bibinfo {pages} {052107} (\bibinfo {year} {2019})}\BibitemShut {NoStop}%
\bibitem [{\citenamefont {Ben~Hayun}\ \emph {et~al.}(2021)\citenamefont
  {Ben~Hayun}, \citenamefont {Reinhardt}, \citenamefont {Nemirovsky},
  \citenamefont {Karnieli}, \citenamefont {Rivera},\ and\ \citenamefont
  {Kaminer}}]{BenHayun2021}%
  \BibitemOpen
  \bibfield  {author} {\bibinfo {author} {\bibfnamefont {A.}~\bibnamefont
  {Ben~Hayun}}, \bibinfo {author} {\bibfnamefont {O.}~\bibnamefont
  {Reinhardt}}, \bibinfo {author} {\bibfnamefont {J.}~\bibnamefont
  {Nemirovsky}}, \bibinfo {author} {\bibfnamefont {A.}~\bibnamefont
  {Karnieli}}, \bibinfo {author} {\bibfnamefont {N.}~\bibnamefont {Rivera}},\
  and\ \bibinfo {author} {\bibfnamefont {I.}~\bibnamefont {Kaminer}},\ }\href
  {https://doi.org/10.1126/sciadv.abe4270} {\bibfield  {journal} {\bibinfo
  {journal} {Science Advances}\ }\textbf {\bibinfo {volume} {7}},\ \bibinfo
  {pages} {eabe4270} (\bibinfo {year} {2021})}\BibitemShut {NoStop}%
\bibitem [{\citenamefont {R{\"a}tzel}\ \emph {et~al.}(2021)\citenamefont
  {R{\"a}tzel}, \citenamefont {Hartley}, \citenamefont {Schwartz},\ and\
  \citenamefont {Haslinger}}]{Ratzel2021}%
  \BibitemOpen
  \bibfield  {author} {\bibinfo {author} {\bibfnamefont {D.}~\bibnamefont
  {R{\"a}tzel}}, \bibinfo {author} {\bibfnamefont {D.}~\bibnamefont {Hartley}},
  \bibinfo {author} {\bibfnamefont {O.}~\bibnamefont {Schwartz}},\ and\
  \bibinfo {author} {\bibfnamefont {P.}~\bibnamefont {Haslinger}},\ }\href
  {https://doi.org/10.1103/PhysRevResearch.3.023247} {\bibfield  {journal}
  {\bibinfo  {journal} {Physical Review Research}\ }\textbf {\bibinfo {volume}
  {3}},\ \bibinfo {pages} {023247} (\bibinfo {year} {2021})}\BibitemShut
  {NoStop}%
\bibitem [{\citenamefont {Di~Giulio}\ \emph {et~al.}(2019)\citenamefont
  {Di~Giulio}, \citenamefont {Kociak},\ and\ \citenamefont {{de
  Abajo}}}]{DiGiulio2019}%
  \BibitemOpen
  \bibfield  {author} {\bibinfo {author} {\bibfnamefont {V.}~\bibnamefont
  {Di~Giulio}}, \bibinfo {author} {\bibfnamefont {M.}~\bibnamefont {Kociak}},\
  and\ \bibinfo {author} {\bibfnamefont {F.~J.~G.}\ \bibnamefont {{de
  Abajo}}},\ }\href {https://doi.org/10.1364/OPTICA.6.001524} {\bibfield
  {journal} {\bibinfo  {journal} {Optica}\ }\textbf {\bibinfo {volume} {6}},\
  \bibinfo {pages} {1524} (\bibinfo {year} {2019})},\ \Eprint
  {https://arxiv.org/abs/1905.06887} {arxiv:1905.06887} \BibitemShut {NoStop}%
\bibitem [{\citenamefont {Tsarev}\ \emph {et~al.}(2021)\citenamefont {Tsarev},
  \citenamefont {Ryabov},\ and\ \citenamefont {Baum}}]{Tsarev2021}%
  \BibitemOpen
  \bibfield  {author} {\bibinfo {author} {\bibfnamefont {M.}~\bibnamefont
  {Tsarev}}, \bibinfo {author} {\bibfnamefont {A.}~\bibnamefont {Ryabov}},\
  and\ \bibinfo {author} {\bibfnamefont {P.}~\bibnamefont {Baum}},\ }\href
  {https://doi.org/10.1103/physrevlett.127.165501} {\bibfield  {journal}
  {\bibinfo  {journal} {Physical Review Letters}\ }\textbf {\bibinfo {volume}
  {127}},\ \bibinfo {pages} {165501} (\bibinfo {year} {2021})}\BibitemShut
  {NoStop}%
\bibitem [{\citenamefont {Kfir}(2019)}]{Kfir2019}%
  \BibitemOpen
  \bibfield  {author} {\bibinfo {author} {\bibfnamefont {O.}~\bibnamefont
  {Kfir}},\ }\href {https://doi.org/10.1103/PhysRevLett.123.103602} {\bibfield
  {journal} {\bibinfo  {journal} {Physical Review Letters}\ }\textbf {\bibinfo
  {volume} {123}},\ \bibinfo {pages} {103602} (\bibinfo {year}
  {2019})}\BibitemShut {NoStop}%
\bibitem [{\citenamefont {Talebi}\ and\ \citenamefont {B{\v
  r}ezinov{\'a}}(2021)}]{Talebi2021}%
  \BibitemOpen
  \bibfield  {author} {\bibinfo {author} {\bibfnamefont {N.}~\bibnamefont
  {Talebi}}\ and\ \bibinfo {author} {\bibfnamefont {I.}~\bibnamefont {B{\v
  r}ezinov{\'a}}},\ }\href {https://doi.org/10.1088/1367-2630/ac06e7}
  {\bibfield  {journal} {\bibinfo  {journal} {New Journal of Physics}\ }\textbf
  {\bibinfo {volume} {23}},\ \bibinfo {pages} {063066} (\bibinfo {year}
  {2021})}\BibitemShut {NoStop}%
\bibitem [{\citenamefont {Kone{\v c}n{\'a}}\ \emph {et~al.}(2022)\citenamefont
  {Kone{\v c}n{\'a}}, \citenamefont {Iyikanat},\ and\ \citenamefont
  {{Garc{\'i}a de Abajo}}}]{Konecna2022}%
  \BibitemOpen
  \bibfield  {author} {\bibinfo {author} {\bibfnamefont {A.}~\bibnamefont
  {Kone{\v c}n{\'a}}}, \bibinfo {author} {\bibfnamefont {F.}~\bibnamefont
  {Iyikanat}},\ and\ \bibinfo {author} {\bibfnamefont {F.~J.}\ \bibnamefont
  {{Garc{\'i}a de Abajo}}},\ }\href {https://doi.org/10.1126/sciadv.abo7853}
  {\bibfield  {journal} {\bibinfo  {journal} {Science Advances}\ }\textbf
  {\bibinfo {volume} {8}},\ \bibinfo {pages} {eabo7853} (\bibinfo {year}
  {2022})}\BibitemShut {NoStop}%
\bibitem [{\citenamefont {Rivera}\ and\ \citenamefont
  {Kaminer}(2020)}]{Rivera2020}%
  \BibitemOpen
  \bibfield  {author} {\bibinfo {author} {\bibfnamefont {N.}~\bibnamefont
  {Rivera}}\ and\ \bibinfo {author} {\bibfnamefont {I.}~\bibnamefont
  {Kaminer}},\ }\href {https://doi.org/10.1038/s42254-020-0224-2} {\bibfield
  {journal} {\bibinfo  {journal} {Nature Reviews Physics}\ }\textbf {\bibinfo
  {volume} {2}},\ \bibinfo {pages} {538} (\bibinfo {year} {2020})}\BibitemShut
  {NoStop}%
\bibitem [{\citenamefont {{Garc{\'i}a de Abajo}}\ and\ \citenamefont
  {Di~Giulio}(2021)}]{GarciadeAbajo2021}%
  \BibitemOpen
  \bibfield  {author} {\bibinfo {author} {\bibfnamefont {F.~J.}\ \bibnamefont
  {{Garc{\'i}a de Abajo}}}\ and\ \bibinfo {author} {\bibfnamefont
  {V.}~\bibnamefont {Di~Giulio}},\ }\href
  {https://doi.org/10.1021/acsphotonics.0c01950} {\bibfield  {journal}
  {\bibinfo  {journal} {ACS Photonics}\ }\textbf {\bibinfo {volume} {8}},\
  \bibinfo {pages} {945} (\bibinfo {year} {2021})}\BibitemShut {NoStop}%
\bibitem [{\citenamefont {Taleb}\ \emph {et~al.}(2023)\citenamefont {Taleb},
  \citenamefont {Hentschel}, \citenamefont {Rossnagel}, \citenamefont
  {Giessen},\ and\ \citenamefont {Talebi}}]{Taleb2023}%
  \BibitemOpen
  \bibfield  {author} {\bibinfo {author} {\bibfnamefont {M.}~\bibnamefont
  {Taleb}}, \bibinfo {author} {\bibfnamefont {M.}~\bibnamefont {Hentschel}},
  \bibinfo {author} {\bibfnamefont {K.}~\bibnamefont {Rossnagel}}, \bibinfo
  {author} {\bibfnamefont {H.}~\bibnamefont {Giessen}},\ and\ \bibinfo {author}
  {\bibfnamefont {N.}~\bibnamefont {Talebi}},\ }\bibfield  {journal} {\bibinfo
  {journal} {Nature Physics}\ }\href
  {https://doi.org/10.1038/s41567-023-01954-3} {10.1038/s41567-023-01954-3}
  (\bibinfo {year} {2023})\BibitemShut {NoStop}%
\bibitem [{\citenamefont {Cook}\ \emph {et~al.}(2009)\citenamefont {Cook},
  \citenamefont {Bronsgeest}, \citenamefont {Hagen},\ and\ \citenamefont
  {Kruit}}]{Cook2009}%
  \BibitemOpen
  \bibfield  {author} {\bibinfo {author} {\bibfnamefont {B.}~\bibnamefont
  {Cook}}, \bibinfo {author} {\bibfnamefont {M.}~\bibnamefont {Bronsgeest}},
  \bibinfo {author} {\bibfnamefont {K.}~\bibnamefont {Hagen}},\ and\ \bibinfo
  {author} {\bibfnamefont {P.}~\bibnamefont {Kruit}},\ }\href
  {https://doi.org/10.1016/j.ultramic.2008.11.024} {\bibfield  {journal}
  {\bibinfo  {journal} {Ultramicroscopy}\ }\textbf {\bibinfo {volume} {109}},\
  \bibinfo {pages} {403} (\bibinfo {year} {2009})}\BibitemShut {NoStop}%
\bibitem [{\citenamefont {Yang}\ \emph {et~al.}(2010)\citenamefont {Yang},
  \citenamefont {Mohammed},\ and\ \citenamefont {Zewail}}]{Yang2010}%
  \BibitemOpen
  \bibfield  {author} {\bibinfo {author} {\bibfnamefont {D.-S.}\ \bibnamefont
  {Yang}}, \bibinfo {author} {\bibfnamefont {O.~F.}\ \bibnamefont {Mohammed}},\
  and\ \bibinfo {author} {\bibfnamefont {A.~H.}\ \bibnamefont {Zewail}},\
  }\href {https://doi.org/10.1073/pnas.1009321107} {\bibfield  {journal}
  {\bibinfo  {journal} {Proceedings of the National Academy of Sciences}\
  }\textbf {\bibinfo {volume} {107}},\ \bibinfo {pages} {14993} (\bibinfo
  {year} {2010})}\BibitemShut {NoStop}%
\bibitem [{\citenamefont {Kuwahara}\ \emph {et~al.}(2016)\citenamefont
  {Kuwahara}, \citenamefont {Nambo}, \citenamefont {Aoki}, \citenamefont
  {Sameshima}, \citenamefont {Jin}, \citenamefont {Ujihara}, \citenamefont
  {Asano}, \citenamefont {Saitoh}, \citenamefont {Takeda},\ and\ \citenamefont
  {Tanaka}}]{Kuwahara2016}%
  \BibitemOpen
  \bibfield  {author} {\bibinfo {author} {\bibfnamefont {M.}~\bibnamefont
  {Kuwahara}}, \bibinfo {author} {\bibfnamefont {Y.}~\bibnamefont {Nambo}},
  \bibinfo {author} {\bibfnamefont {K.}~\bibnamefont {Aoki}}, \bibinfo {author}
  {\bibfnamefont {K.}~\bibnamefont {Sameshima}}, \bibinfo {author}
  {\bibfnamefont {X.}~\bibnamefont {Jin}}, \bibinfo {author} {\bibfnamefont
  {T.}~\bibnamefont {Ujihara}}, \bibinfo {author} {\bibfnamefont
  {H.}~\bibnamefont {Asano}}, \bibinfo {author} {\bibfnamefont
  {K.}~\bibnamefont {Saitoh}}, \bibinfo {author} {\bibfnamefont
  {Y.}~\bibnamefont {Takeda}},\ and\ \bibinfo {author} {\bibfnamefont
  {N.}~\bibnamefont {Tanaka}},\ }\href {https://doi.org/10/ggnzpb} {\bibfield
  {journal} {\bibinfo  {journal} {Applied Physics Letters}\ }\textbf {\bibinfo
  {volume} {109}},\ \bibinfo {pages} {013108} (\bibinfo {year}
  {2016})}\BibitemShut {NoStop}%
\bibitem [{\citenamefont {Bronsgeest}\ \emph {et~al.}(2007)\citenamefont
  {Bronsgeest}, \citenamefont {Barth}, \citenamefont {Schwind}, \citenamefont
  {Swanson},\ and\ \citenamefont {Kruit}}]{Bronsgeest2007}%
  \BibitemOpen
  \bibfield  {author} {\bibinfo {author} {\bibfnamefont {M.~S.}\ \bibnamefont
  {Bronsgeest}}, \bibinfo {author} {\bibfnamefont {J.~E.}\ \bibnamefont
  {Barth}}, \bibinfo {author} {\bibfnamefont {G.~A.}\ \bibnamefont {Schwind}},
  \bibinfo {author} {\bibfnamefont {L.~W.}\ \bibnamefont {Swanson}},\ and\
  \bibinfo {author} {\bibfnamefont {P.}~\bibnamefont {Kruit}},\ }\href
  {https://doi.org/10.1116/1.2794067} {\bibfield  {journal} {\bibinfo
  {journal} {Journal of Vacuum Science \& Technology B: Microelectronics and
  Nanometer Structures}\ }\textbf {\bibinfo {volume} {25}},\ \bibinfo {pages}
  {2049} (\bibinfo {year} {2007})}\BibitemShut {NoStop}%
\bibitem [{\citenamefont {Kruit}\ and\ \citenamefont
  {Jansen}(2009)}]{Kruit2009}%
  \BibitemOpen
  \bibfield  {author} {\bibinfo {author} {\bibfnamefont {P.}~\bibnamefont
  {Kruit}}\ and\ \bibinfo {author} {\bibfnamefont {G.~H.}\ \bibnamefont
  {Jansen}},\ }in\ \href@noop {} {\emph {\bibinfo {booktitle} {Handbook of
  Charged Particle Optics}}}\ (\bibinfo  {publisher} {{CRC Press}},\ \bibinfo
  {address} {{Boca Raton}},\ \bibinfo {year} {2009})\ \bibinfo {edition} {2nd}\
  ed.,\ pp.\ \bibinfo {pages} {275--318}\BibitemShut {NoStop}%
\bibitem [{\citenamefont {Jansen}(1988)}]{Jansen1988_thesis}%
  \BibitemOpen
  \bibfield  {author} {\bibinfo {author} {\bibfnamefont {G.~H.}\ \bibnamefont
  {Jansen}},\ }\emph {\bibinfo {title} {Coulomb Interactions in Particle
  Beams}},\ \href@noop {} {\bibinfo {type} {Doctoral thesis}},\ \bibinfo
  {school} {Delft University of Technology}, \bibinfo {address} {{Delft}}
  (\bibinfo {year} {1988})\BibitemShut {NoStop}%
\bibitem [{\citenamefont {Alhassid}(2000)}]{Alhassid2000}%
  \BibitemOpen
  \bibfield  {author} {\bibinfo {author} {\bibfnamefont {Y.}~\bibnamefont
  {Alhassid}},\ }\href {https://doi.org/10.1103/RevModPhys.72.895} {\bibfield
  {journal} {\bibinfo  {journal} {Reviews of Modern Physics}\ }\textbf
  {\bibinfo {volume} {72}},\ \bibinfo {pages} {895} (\bibinfo {year}
  {2000})}\BibitemShut {NoStop}%
\bibitem [{\citenamefont {VandenBussche}\ and\ \citenamefont
  {Flannigan}(2019)}]{VandenBussche2019}%
  \BibitemOpen
  \bibfield  {author} {\bibinfo {author} {\bibfnamefont {E.~J.}\ \bibnamefont
  {VandenBussche}}\ and\ \bibinfo {author} {\bibfnamefont {D.~J.}\ \bibnamefont
  {Flannigan}},\ }\href {https://doi.org/10.1021/acs.nanolett.9b03074}
  {\bibfield  {journal} {\bibinfo  {journal} {Nano Letters}\ }\textbf {\bibinfo
  {volume} {19}},\ \bibinfo {pages} {6687} (\bibinfo {year}
  {2019})}\BibitemShut {NoStop}%
\bibitem [{\citenamefont {Kisielowski}\ \emph {et~al.}(2019)\citenamefont
  {Kisielowski}, \citenamefont {Specht}, \citenamefont {Freitag}, \citenamefont
  {Kieft}, \citenamefont {Verhoeven}, \citenamefont {Rens}, \citenamefont
  {Mutsaers}, \citenamefont {Luiten}, \citenamefont {Rozeveld}, \citenamefont
  {Kang}, \citenamefont {McKenna}, \citenamefont {Nickias},\ and\ \citenamefont
  {Yancey}}]{Kisielowski2019}%
  \BibitemOpen
  \bibfield  {author} {\bibinfo {author} {\bibfnamefont {C.}~\bibnamefont
  {Kisielowski}}, \bibinfo {author} {\bibfnamefont {P.}~\bibnamefont {Specht}},
  \bibinfo {author} {\bibfnamefont {B.}~\bibnamefont {Freitag}}, \bibinfo
  {author} {\bibfnamefont {E.~R.}\ \bibnamefont {Kieft}}, \bibinfo {author}
  {\bibfnamefont {W.}~\bibnamefont {Verhoeven}}, \bibinfo {author}
  {\bibfnamefont {J.~F.~M.}\ \bibnamefont {Rens}}, \bibinfo {author}
  {\bibfnamefont {P.}~\bibnamefont {Mutsaers}}, \bibinfo {author}
  {\bibfnamefont {J.}~\bibnamefont {Luiten}}, \bibinfo {author} {\bibfnamefont
  {S.}~\bibnamefont {Rozeveld}}, \bibinfo {author} {\bibfnamefont
  {J.}~\bibnamefont {Kang}}, \bibinfo {author} {\bibfnamefont {A.~J.}\
  \bibnamefont {McKenna}}, \bibinfo {author} {\bibfnamefont {P.}~\bibnamefont
  {Nickias}},\ and\ \bibinfo {author} {\bibfnamefont {D.~F.}\ \bibnamefont
  {Yancey}},\ }\href {https://doi.org/10.1002/adfm.201807818} {\bibfield
  {journal} {\bibinfo  {journal} {Advanced Functional Materials}\ }\textbf
  {\bibinfo {volume} {29}},\ \bibinfo {pages} {1807818} (\bibinfo {year}
  {2019})}\BibitemShut {NoStop}%
\bibitem [{\citenamefont {Krivanek}\ \emph {et~al.}(2009)\citenamefont
  {Krivanek}, \citenamefont {Ursin}, \citenamefont {Bacon}, \citenamefont
  {Corbin}, \citenamefont {Dellby}, \citenamefont {Hrncirik}, \citenamefont
  {Murfitt}, \citenamefont {Own},\ and\ \citenamefont
  {Szilagyi}}]{Krivanek2009}%
  \BibitemOpen
  \bibfield  {author} {\bibinfo {author} {\bibfnamefont {O.~L.}\ \bibnamefont
  {Krivanek}}, \bibinfo {author} {\bibfnamefont {J.~P.}\ \bibnamefont {Ursin}},
  \bibinfo {author} {\bibfnamefont {N.~J.}\ \bibnamefont {Bacon}}, \bibinfo
  {author} {\bibfnamefont {G.~J.}\ \bibnamefont {Corbin}}, \bibinfo {author}
  {\bibfnamefont {N.}~\bibnamefont {Dellby}}, \bibinfo {author} {\bibfnamefont
  {P.}~\bibnamefont {Hrncirik}}, \bibinfo {author} {\bibfnamefont {M.~F.}\
  \bibnamefont {Murfitt}}, \bibinfo {author} {\bibfnamefont {C.~S.}\
  \bibnamefont {Own}},\ and\ \bibinfo {author} {\bibfnamefont {Z.~S.}\
  \bibnamefont {Szilagyi}},\ }\href {https://doi.org/10.1098/rsta.2009.0087}
  {\bibfield  {journal} {\bibinfo  {journal} {Philosophical Transactions of the
  Royal Society A: Mathematical, Physical and Engineering Sciences}\ }\textbf
  {\bibinfo {volume} {367}},\ \bibinfo {pages} {3683} (\bibinfo {year}
  {2009})}\BibitemShut {NoStop}%
\bibitem [{\citenamefont {Meier}\ \emph {et~al.}(2022)\citenamefont {Meier},
  \citenamefont {Heimerl},\ and\ \citenamefont {Hommelhoff}}]{Meier2022}%
  \BibitemOpen
  \bibfield  {author} {\bibinfo {author} {\bibfnamefont {S.}~\bibnamefont
  {Meier}}, \bibinfo {author} {\bibfnamefont {J.}~\bibnamefont {Heimerl}},\
  and\ \bibinfo {author} {\bibfnamefont {P.}~\bibnamefont {Hommelhoff}},\
  }\href@noop {} {\bibinfo {title} {Few-electron correlations after ultrafast
  photoemission from nanometric needle tips}} (\bibinfo {year} {2022}),\
  \Eprint {https://arxiv.org/abs/2209.11806} {arxiv:2209.11806 [cond-mat]}
  \BibitemShut {NoStop}%
\bibitem [{\citenamefont {{van Schayck}}\ \emph {et~al.}(2020)\citenamefont
  {{van Schayck}}, \citenamefont {{van Genderen}}, \citenamefont {Maddox},
  \citenamefont {Roussel}, \citenamefont {Boulanger}, \citenamefont
  {Fr{\"o}jdh}, \citenamefont {Abrahams}, \citenamefont {Peters},\ and\
  \citenamefont {Ravelli}}]{vanSchayck2020}%
  \BibitemOpen
  \bibfield  {author} {\bibinfo {author} {\bibfnamefont {J.~P.}\ \bibnamefont
  {{van Schayck}}}, \bibinfo {author} {\bibfnamefont {E.}~\bibnamefont {{van
  Genderen}}}, \bibinfo {author} {\bibfnamefont {E.}~\bibnamefont {Maddox}},
  \bibinfo {author} {\bibfnamefont {L.}~\bibnamefont {Roussel}}, \bibinfo
  {author} {\bibfnamefont {H.}~\bibnamefont {Boulanger}}, \bibinfo {author}
  {\bibfnamefont {E.}~\bibnamefont {Fr{\"o}jdh}}, \bibinfo {author}
  {\bibfnamefont {J.-P.}\ \bibnamefont {Abrahams}}, \bibinfo {author}
  {\bibfnamefont {P.~J.}\ \bibnamefont {Peters}},\ and\ \bibinfo {author}
  {\bibfnamefont {R.~B.}\ \bibnamefont {Ravelli}},\ }\href
  {https://doi.org/10.1016/j.ultramic.2020.113091} {\bibfield  {journal}
  {\bibinfo  {journal} {Ultramicroscopy}\ }\textbf {\bibinfo {volume} {218}},\
  \bibinfo {pages} {113091} (\bibinfo {year} {2020})}\BibitemShut {NoStop}%
\bibitem [{\citenamefont {{van Schayck, J.
  Paul}}(2021)}]{vanSchayckJ.Paul2021}%
  \BibitemOpen
  \bibfield  {author} {\bibinfo {author} {\bibnamefont {{van Schayck, J.
  Paul}}},\ }\href {https://doi.org/10.5281/ZENODO.4580458} {\bibinfo {title}
  {{{M4I-nanoscopy}}/{{tpx3HitParser}}: {{Version}} 2.1.0}},\ \bibinfo
  {howpublished} {Zenodo} (\bibinfo {year} {2021})\BibitemShut {NoStop}%
\end{thebibliography}%
\bibliographystyle{apsrev4-2}
\end{document}